 \documentclass[11pt]{article}

\usepackage{amsmath,amsfonts,amssymb}
 \usepackage[numbers,sort&compress]{natbib}

 \usepackage{braket}
 \usepackage{dsfont}
 
 \usepackage{hyperref}
\hypersetup{
   colorlinks   =  true,
    linkcolor    = blue,
    citecolor    = red,
     urlcolor	=blue,
}

\numberwithin{equation}{section}

\newtheorem{rem}{Remark} 

\newcommand\be{\begin{equation}}
\newcommand\ee{\end{equation}}
\newcommand\bea{\begin{eqnarray}}
\newcommand\eea{\end{eqnarray}}

  \usepackage{color}

\newcommand{\JJ}{J} 
\newcommand{\CC}{\mathcal{C}} 
  \newcommand{\LL}{L} 

  \newcommand{\OO}{{\mathcal{O}}} 

\newcommand{\ff}{\Psi} 

\begin{document}

 \noindent
 {\Large \bf
The Dunkl two-photon/Schr{\"o}dinger algebras and their applications}

\medskip

\medskip

\begin{center}
 
{\sc    Francisco J.~Herranz$^{1,\dagger}$ and  Danilo Latini$^{2,3,\ddagger}$ 
}

 \end{center}

\medskip

\noindent
{$^1$ Departamento de F\'isica, Universidad de Burgos,
E-09001 Burgos, Spain}

\noindent
{$^2$ Dipartimento di Matematica ``Federigo Enriques", Universit\`a degli Studi di Milano, Via C. Saldini 50, 20133 Milano, Italy}

\noindent
{$^3$  INFN Sezione di Milano, Via G. Celoria 16, 20133 Milano, Italy}

  \medskip

\noindent   
{ {\footnotesize
  $^\dagger$\href{mailto:fjherranz@ubu.es}{\texttt{fjherranz@ubu.es}}\ \   $^\ddagger$\href{mailto:danilo.latini@unimi.it}{\texttt{danilo.latini@unimi.it}} 
 }
 }

\begin{abstract}
\noindent
The work introduces the novel Dunkl two-photon and Dunkl--Schr{\"o}dinger algebras as reflection-extended counterparts of the standard six-dimensional two-photon and  Schr{\"o}dinger algebras. These extensions are realized through the presence of an involutive generator representing a reflection, and they reduce to the undeformed two-photon and Schr{\"o}dinger algebras in the absence of this additional generator in an appropriate limit. For both algebraic structures, which are shown to be isomorphic, we derive the associated cubic polynomial Casimir invariant. Furthermore, we identify several structurally significant subalgebras and compute their corresponding quadratic Casimir invariants. Representations of the Dunkl two-photon algebra are constructed on the Fock space of the one-dimensional Dunkl harmonic oscillator and subsequently translated into holomorphic realizations within the associated Dunkl--Fock--Bargmann representation. Finally, for the Dunkl--Schr{\"o}dinger algebra, we present an explicit vector field realization in terms of the  Dunkl spatial derivative, while keeping  time continuous. When applied to the Casimir invariants of the aforementioned subalgebras, this realization yields a family of six nontrivial Dunkl differential-difference equations in $1+1$ dimensions, including, in particular, a Dunkl version of the well-known heat-Schr{\"o}dinger equation.
\end{abstract}

\bigskip

\noindent
\textbf{Keywords}: {two-photon algebra; Schr\"{o}dinger algebra;  Dunkl operators;  Casimir invariants;\\  representations;  Dunkl harmonic oscillator;  differential-difference equations}

\medskip

\bigskip

\newpage

\tableofcontents


\section{Introduction}
\label{intro}

The so-called two-photon algebra  $\mathfrak{h}_6$ is a non-semisimple six-dimensional Lie algebra, which was formally introduced in~\cite{ZFG},  and  emerged naturally within the framework of quantum optics, more precisely in relation to squeezed states~{\cite{Brif}}.  In particular, let us consider the time-dependent Hamiltonian given by
\be
\mathcal{H}= \hbar \omega \big(\hat{a}^\dagger \hat{a}  +\tfrac 12 \big) + f_1(t) \hat{a}^\dagger +   f_1^\ast(t) \hat{a} + f_2(t) \big(\hat{a}^\dagger\big)^2 +   f_2^\ast(t) \hat{a}^2 \, ,
\label{a1}
\ee
where $f_1$ and $f_2$  are  time-dependent complex functions, and $\hat{a}$, $\hat{a}^\dagger$ are the standard annihilation/creation boson operators that generate the Heisenberg--Weyl algebra, such that $[\hat{a},\hat{a}^\dagger]=\hat{\mathds{1}}$, with  $\hat{\mathds{1}}$ being the identity operator.
This Hamiltonian corresponds to a single-mode radiation field describing the generation of a squeezed coherent state, in which  the processes of coherent state formation and squeezing are determined by the time dependence of the functions $f_1$ and $f_2$.  According to~\cite[\S V.B.1 p.~895]{ZFG}, the operators underlying $\mathcal{H}$, together with the identity, span the  two-photon algebra:
\be
\mathfrak{h}_6={\rm span} \big\{ \hat{a}^\dagger \hat{a}  +\tfrac 12,  \hat{a}^\dagger,\hat{a} , \big(\hat{a}^\dagger\big)^2,    \hat{a}^2,   \hat{\mathds{1}} \big\}\, .
\label{a2}
\ee
Therefore, one-photon processes are provided by the oscillator subalgebra 
$$
\mathfrak{h}_4={\rm span} \big\{ \hat{a}^\dagger \hat{a}  +\tfrac 12,  \hat{a}^\dagger,\hat{a}  ,      \hat{\mathds{1}} \big\}\subset \mathfrak{h}_6\, ,
$$
while single-mode two-photon processes   are determined by the subalgebra
$$
\mathfrak{su}(1,1)={\rm span} \big\{ \hat{a}^\dagger \hat{a}  +\tfrac 12,   \big(\hat{a}^\dagger\big)^2,    \hat{a}^2 \big\}\subset \mathfrak{h}_6\, .
$$
Given the physical significance of  $\mathfrak{h}_6$ and following the pioneering works~\cite{ZFG,Brif}, 
the two-photon algebra has been studied in various contexts, which we describe below.

Two Hopf algebra deformations for $\mathfrak{h}_6$  were constructed in~\cite{PBH1997} and~\cite{PBH1998} from the quantum group formalism~\cite{CP94,Abe2004}, both giving rise to non-equivalent ``deformed" boson representations of the Hamiltonian $\mathcal{H}$ (\ref{a1}). The former in~\cite{PBH1997} was based in the one-photon operator $\hat{a}^\dagger $, while the latter in~\cite{PBH1998}  made use of the  two-photon operator $\big(\hat{a}^\dagger\big)^2$. 
In both cases, the quantum deformation parameter was shown to provide deformed states of light, which could be understood as some kind of ``perturbations" of the initial coherent/squeezed states in~\cite{ZFG,Brif}. Regarding the quantum deformations of  $\mathfrak{h}_6$, all its Lie bialgebras can be found in~\cite{PBH1999}; these are,  in fact, the structures which arise at the  first-order of the  quantum deformation parameter. In addition,  the  two-photon algebra  $\mathfrak{h}_6$  has also been used within the framework of classical integrable systems~\cite{Perelomov}  as a Lie--Poisson algebra~\cite{BBH09}, where the classical counterpart of the quantum operators (\ref{a2}) is expressed in terms of  the canonical position $q$ and momentum  $p$ variables, with Poisson bracket $\{q,p\}=1$.

Recently, new findings and applications regarding $\mathfrak{h}_6$ have been obtained.
The notion of coalgebra symmetry (see~\cite{BBHMR2009} and references therein)  for discrete-time systems and  the role of the Lie--Poisson algebra $\mathfrak{h}_6$ have been discussed in~\cite{GLT2023}. Subsequently, a more detailed study devoted to the two-photon algebra has been presented  in \cite{DG2025}, in which  all discrete-time systems in quasi-standard form admitting coalgebra symmetry with respect to the Lie--Poisson algebra $\mathfrak{h}_6$ have been characterized. The two-photon algebra has also been used in the 
context of Lie--Hamilton systems~\cite{LS20}, {\em i.e.}, Lie systems with a compatible symplectic structure, in~\cite{CCH2025}, as a subalgebra of the symplectic Lie algebra $\mathfrak{sp}(4,\mathbb R)$.
Finally, in~\cite{GLG2026}, a novel chain of Lie algebras, denoted by $\mathfrak{g}_n$ for $n\geq 2$, whose $n=3$ member is isomorphic to  $\mathfrak{h}_6$,  has been introduced and  the hierarchy of integrable systems arising from the corresponding coalgebra symmetry has been studied.

 It is worth noting that in~\cite{PBH1997}  it was shown that  the two-photon algebra is isomorphic to the  centrally extended Schr\"odinger algebra in $1+1$ dimensions $\mathcal{S}$~\cite{Hagen,Niederer,Feinsilver2004}, which  is the Lie algebra of the symmetry group of the $(1 + 1)$-dimensional  heat-Schr\"odinger equation. This property of $\mathcal{S}$ is obtained through 
a differential  realization of the Schr\"odinger generators  in terms of the time and space coordinates $(t, x)$. In this respect, we recall that the two particular Hopf algebra deformations for $\mathfrak{h}_6$    mentioned above  were  algebraically ``translated" into quantum Schr\"odinger algebras 
and, subsequently, by means of a suitable  differential-difference realization, governed by the quantum deformation parameter, a time discretization of the heat-Schr\"odinger equation was derived in~\cite{PBH1998}, while a discrete space Schr\"odinger equation was constructed in~\cite{BHNN2000}. All Schr\"odinger Lie bialgebras and their Poisson--Lie groups can be found in~\cite{BHP2000}.

The aim of this work is twofold. Firstly, we construct the Dunkl  two-photon algebra, which 
    incorporates the presence of differential-difference operators introduced by C.~F.~Dunkl in~\cite{Dunkl1989}. These operators are associated with finite reflection groups \cite{hump1990} and can be regarded as a generalization of the standard derivatives \cite{Ros2003, Dunkl2012}.  
    Secondly, we derive  the Dunkl Schr\"odinger algebra. Furthermore, we present in detail some  applications of these two novel Dunkl algebras.
 
More specifically, the structure of the article is as follows. In the next section, we summarize the fundamentals  of the two-photon algebra, covering its algebraic structures and relevant physical  realizations, which we will use throughout the paper. In section~\ref{sec3}, we construct the  Dunkl  two-photon algebra and also obtain some compatible algebraic structures, such as the Casimir invariant and a description of the main Dunkl two-photon subalgebras. Next, we introduce  the Dunkl--Fock space and    Dunkl--Fock--Bargmann representations, which are applied to the construction  of the  Dunkl harmonic oscillator from the Dunkl  two-photon algebra (for isotropic Dunkl  oscillators, see~\cite{
GIVZ2013I,GIVZ2013II,GVZ2014,Ghaz2021,HL2025} and references therein). In fact, the representation Hilbert space of the Dunkl two-photon algebra is the Fock space of the Dunkl harmonic oscillator. This goes in parallel to the standard two-photon Lie algebra for which the representation Hilbert space is given by the Fock space of the quantum harmonic oscillator~{\cite{Brif}}. In section~\ref{sec4},
we present the main features of the centrally extended   Schr\"odinger algebra in $1+1$ dimensions, 
establishing the notation used in this work and its role as the  Lie symmetry algebra of the  heat-Schr\"odinger equation. Then, in section~\ref{sec5}, we derive the  Dunkl--Schr\"odinger algebra, study its main  algebraic structures,  including the cubic Casimir invariant and several notable subalgebras, and apply them to the construction of 
six new Dunkl differential-difference equations in $1 + 1$ dimensions, including, among others, the Dunkl counterpart of the well-known heat-Schr\"odinger equation. Finally, some remarks and open perspectives close the paper.


\section{The two-photon Lie algebra}
\label{sec2}

The  two-photon  algebra $\mathfrak{h}_6$~\cite{ZFG, Brif}  is a six-dimensional Lie algebra spanned by the generators 
$\{N, A_+, A_-, B_+, B_-, M\}$, which satisfy the following defining commutation relations~\cite{PBH1997}:
\begin{equation}
\begin{aligned}
{[N, A_+] }&= A_+ \, , \qquad\quad\quad\!  [N, A_-]  = -A_- \, , &\quad  [N, B_+] &= 2B_+ \, , \\[2pt]
[N, B_-] &= -2B_- \, ,\qquad [A_-, B_+]  = 2A_+  \, ,&\quad  [A_-, A_+] &= M   \, ,  \\[2pt]
 [A_+, B_+] &= 0 \, , \qquad\qquad\,  [A_+, B_-]  = -2A_-    \, , &\quad 	[A_-, B_-] &= 0  \, ,  \\[2pt]
[B_-, B_+] &= 4N+2 M  \, ,  \qquad\!	 [M, \cdot\, ]  = 0  \, .   &&
\label{Two-Photon}
\end{aligned}
\end{equation}
Hence, $M$ is a central nontrivial generator and $\mathfrak{h}_6$ contains, among others,  some relevant Lie subalgebras:
\begin{itemize}
\item The Heisenberg--Weyl algebra  $\mathfrak{h}_3={\rm span}\{ A_+,A_-,M \}$.
\item The oscillator algebra  $\mathfrak{h}_4={\rm span}\{ N,A_+,A_-,M \}$.
\item $\mathfrak{gl}(2)={\rm span}\{N, B_+,B_-,M \}$.
\end{itemize}
  Thus, we have the subalgebra embeddings:
\begin{equation}
\mathfrak{h}_3\subset \mathfrak{h}_4 \subset \mathfrak{h}_6\,  ,\qquad
\mathfrak{gl}(2)  \subset \mathfrak{h}_6\,  .
\label{embed}
\end{equation}

The two-photon algebra is endowed with the automorphism defined by
\be
\{N, A_+, A_-, B_+, B_-, M\} \to \{-N, -A_-, -A_+, -B_-, -B_+, -M\} \, ,
\label{aut0}
\ee
which preserves the commutation rules (\ref{Two-Photon}) and also holds for its subalgebras. 
 
Besides the central generator $M$, the Lie algebra $\mathfrak{h}_6$ has a third-order Casimir, which can be deduced from those corresponding to the so-called   optical algebra~\cite{Burdet1978}, namely
\begin{align}
\CC_{\mathfrak{h}_6} &=B_+ B_- M- A_-^2 B_+- A_+^2 B_- - N^2 M -  N M^2 + 2 N A_+ A_- \nonumber\\
&\qquad +  A_+ A_- M + 3 N M   \, ,
	\label{eq:Casimir2}
\end{align}
such that $\big[\CC_{\mathfrak{h}_6} , X \big] =0$ for all $X\in{\mathfrak{h}_6} $.

Applications of  $\mathfrak{h}_6$  in quantum optics~\cite{ZFG, Brif} are provided through its representations. In fact, the name  ``{\em two-photon algebra}" means that $\mathfrak{h}_6$ can
  be used to generate a large amount of squeezed and coherent states for (single mode) one- and two-photon processes under the representation given by
\begin{equation}
\begin{aligned}
	\hat{N}&=\hat{a}^\dagger \hat{a}  \, ,\quad  &\hat{A}_+ &=\hat{a}^\dagger \, , \quad  &\hat{A}_-&=\hat{a}\, ,\\
	\hat{M}&=\hat{\mathds{1}}\, ,\quad\ &\hat{B}_+&=\big(\hat{a}^\dagger\big)^2 \, , \quad  &\hat{B}_-&=\hat{a}^2 \, ,
\end{aligned}
 \label{eq:reps0}
 \end{equation}
 such that
 \be
\big[ \hat{a} , \hat{a}^\dagger \big]=\hat{\mathds{1}}\, ,
 \label{h3}
 \ee
where   $\hat{\mathds{1}}$ denotes the identity operator and it is assumed that $\hbar=1$. 
  These operators satisfy the Lie brackets (\ref{Two-Photon}) and act 
on the number-state  Hilbert space spanned by the orthonormal states $\{\ket{n} \}_{n=0}^{\infty}$, {\em i.e.}, the Fock space~\cite{Fock},  as follows
 \begin{equation}
\begin{aligned}
\hat{N} \ket{n}&=n \ket{n}\, ,\quad 	&\hat{M} \ket{n}&=\ket{n} \, ,  \\[2pt]
\hat{A}_+ \ket{n}&=\sqrt{n+1}\ket{n+1}\, ,\quad 	&\hat{A}_- \ket{n}&=\sqrt{n}\ket{n-1} \, ,  \\[2pt]
	\hat{B}_+ \ket{n}&=\sqrt{(n+1) (n+2)}\ket{n+2} \, ,\quad 	&\hat{B}_- \ket{n}&=\sqrt{n(n-1)}\ket{n-2}  \, .
\label{eq:action0}
\end{aligned}
\end{equation}
These relations explicitly show the role of the generators $\hat{N}$, $\hat{A}_\pm$ and $\hat{B}_\pm$ as   number, 
(creation and annihilation) one-photon and two-photon operators, respectively. Therefore,  one-photon processes are algebraically encoded within the oscillator subalgebra $\mathfrak{h}_4$, while the subalgebra  $\mathfrak{gl}(2)  $  contains the information concerning two-photon dynamics.

In addition, the representation (\ref{eq:action0}) can be translated into a Fock--Bargmann representation~\cite{Bargmann}, where the $\mathfrak{h}_6$-generators   act in the Hilbert space of entire analytic functions $f(z)$ on ${\mathbb{C}}$ as linear differential operators; these are
\begin{equation}
\begin{aligned}
\bar{N} &= z\, \frac{\text{d}}{\text{d}z}  \, ,
\quad  &\bar{A}_+ &= z\, ,  \quad 
&\bar{A}_- &=  \frac{\text{d}}{\text{d}z}\, , \\[2pt]
\bar{M} &= \bar{\mathds{1}} \, ,  \quad &\bar{B}_+ &= z^2\, , \quad 
&\bar{B}_- &=  \frac{\text{d}^2}{\text{d}z^2}\   , 
\end{aligned}
\label{eq:reps2z0}
\end{equation}
where   $\bar{\mathds{1}}$ is the identity operator. Note that, in both representations (\ref{eq:reps0}) and (\ref{eq:reps2z0}), the central generator $M$   is always the identity operator: $M\equiv\hat{\mathds{1}}\equiv \bar{\mathds{1}}$. Then,  the two-photon algebra eigenstates~\cite{Brif,PBH1998} are determined by the analytic eigenfunctions that satisfy 
\begin{equation}
\big(\beta_1 \bar{N}+ \beta_2 \bar{B}_-+\beta_3 \bar{B}_++\beta_4 \bar{A}_-+\beta_5 \bar{A}_+ \big)f(z)=\lambda f(z) \, ,
\label{eq:eigentwophoton0}
\end{equation}
where all $\beta_i$'s   are arbitrary complex coefficients and $\lambda$ is a complex eigenvalue.
By introducing (\ref{eq:reps2z0}), we arrive at the following differential equation
 \begin{equation}
	\beta_2 f''(z)+(\beta_1 z+\beta_4) f'(z)+\big(\beta_3 z^2+\beta_5 z-\lambda \big)f(z)=0 \, .
	\label{eq:defeqmu01}
\end{equation}
The solutions of this  $\mathfrak{h}_6$-equation leads to the two-photon coherent/squeezed states~\cite{Brif}. Particular one- and two-photon coherent/squeezed states  are deduced from 
 the subalgebras $\mathfrak{h}_4$ and $\mathfrak{gl}(2)$   by setting $\beta_2 = \beta_3 = 0$ and $\beta_4 = \beta_5 = 0$, respectively.
 

\section{The Dunkl two-photon algebra}
\label{sec3}

In this section, we focus on the first aim of the paper:   to construct and develop the novel seven-dimensional  Dunkl two-photon algebra, denoted by $\mathfrak{h}_6^{(\alpha)}$ with parameter $\alpha \in \mathbb{R}$, which extends the  two-photon Lie algebra $\mathfrak{h}_6$~\cite{ZFG,Brif}, briefly described in the previous section. This approach requires the inclusion of a reflection generator $R$, namely an involutive generator satisfying $R^2=\mathrm{Id}$, where $\mathrm{Id}$ is the identity generator and $\alpha$ is the  parameter associated with $R$.

Specifically, the defining relations and structural properties of $\mathfrak{h}_6^{(\alpha)}$, together with its associated cubic Casimir invariant,    are presented in section~\ref{subsec3.1}. Six distinguished subalgebras of $\mathfrak{h}_6^{(\alpha)}$ are then discussed, along with their corresponding quadratic Casimir invariants, and several nontrivial relations among these Casimir polynomials are also derived in section~\ref{subsec3.2}. Finally,  useful representations of this algebraic structure and their applications are developed, including a Dunkl--Fock space representation  and a Dunkl--Fock--Bargmann realization   in sections~\ref{subsec3.3} and~\ref{subsec3.4}, respectively.  In particular, the applications  studied here include the Dunkl harmonic oscillator and the differential-difference equations that extend the standard differential equation (\ref{eq:defeqmu01}), providing two-photon algebra eigenstates, to the Dunkl framework.


\subsection{Definition, structural properties and Casimir invariant of $\mathfrak{h}_6^{(\alpha)}$}
\label{subsec3.1}

Let us consider the seven generators $\{N, A_+, A_-, B_+, B_-, M, R\}$ ordered as follows:
$$N  \prec A_+  \prec A_-  \prec B_+  \prec B_-  \prec M  \prec R$$
and a parameter $\alpha \in \mathbb{R}$.  Let $\mathfrak{h}_6^{(\alpha)}$ be the  unital associative algebra generated by
\begin{equation}
	\{ N,\, A_+,\, A_-,\, B_+,\, B_-,\, M,\, R \}\, ,
	\label{generators}
\end{equation}
with associative product (denoted by juxtaposition) and identity element $\mathrm{Id}$. 

 The {\em Dunkl two-photon algebra}  $\mathfrak{h}_6^{(\alpha)}$ is determined by the following defining commutation relations
\begin{equation}
\begin{aligned}
	{[N, A_+] }&= A_+ \, , \qquad\quad\quad\! 
	 [N, A_-]  = -A_- \, ,
	&  [N, B_+] &= 2B_+ \, ,
	  \\[2pt]
	[N, B_-] &= -2B_- \, ,\qquad
	   [A_-, B_+]  = 2A_+  \, ,
	&  [A_-, A_+] &= M (\mathrm{Id} + \alpha R) \, ,  \\[2pt]
	   [A_+, B_+]  &= 0 \, , \qquad\qquad\, 
	    [A_+, B_-]  = -2A_-    \, , & 	[A_-, B_-] &= 0  \, ,  \\[2pt]
	 [B_-, B_+] &= 4N+2(1+\alpha)M  \, ,  \qquad	 [M, \cdot \, ]  = 0  \, ,   &&
	\label{Dunkl Two-Photon}
\end{aligned}
\end{equation}
along with the additional ones involving the additional generator $R$, namely
\begin{equation}
	R^2 = \mathrm{Id}\, , \qquad  R N =  N R \, , \qquad   R A_\pm=-A_\pm R \, , \qquad 
	R B_\pm  = B_\pm R \, , \qquad   
	R M = M R \, .
	\label{addrel}
\end{equation}
The $\mathfrak{h}_6$-automorphism defined by (\ref{aut0}) is extended to the  Dunkl two-photon algebra in the form
\be
\{N, A_+, A_-, B_+, B_-, M, R\} \to \{-N, -A_-, -A_+, -B_-, -B_+, -M,R\} \, ,
\nonumber
\ee
thus keeping its defining   commutation relations given by (\ref{Dunkl Two-Photon}) and (\ref{addrel}).

   The generator $R$ induces a $\mathbb{Z}_2$ grading of the underlying vector space:
\begin{equation*}
	\mathfrak{h}_6^{(\alpha)}= \mathfrak{h}_{6,0}^{(\alpha)} \oplus \mathfrak{h}_{6,1}^{(\alpha)}\, ,
\end{equation*}
where
\begin{equation*}
	\mathfrak{h}_{6,0}^{(\alpha)}:= \mathrm{span}\{ N, B_\pm, M, R \} \, ,
	\qquad 
	\mathfrak{h}_{6,1}^{(\alpha)}:= \mathrm{span}\{ A_\pm \} \, ,
\end{equation*}
and
\begin{equation*}
	R X = (-1)^{|X|} X R \, ,
	\quad \text{with}\quad |X| = 
	\begin{cases}
		0\, , & X \in \mathfrak{h}_{6,0}^{(\alpha)} \, ,\\[4pt]
		1\, , & X \in \mathfrak{h}_{6,1}^{(\alpha)} \, .
	\end{cases}
\end{equation*}
Thus $R$ acts as a reflection generator distinguishing even and odd generators.

\begin{rem}
	The structure $\big(\mathfrak{h}_6^{(\alpha)}, \boldsymbol{\cdot}\, , \mathrm{Id}\big)$ is an associative unital noncommutative algebra. The commutator
\begin{equation*}
		[X,Y] := XY - YX \, , \qquad X,Y \in \mathfrak{h}_6^{(\alpha)} \, ,
\end{equation*}
endows $\mathfrak{h}_6^{(\alpha)}$ with the structure of a generally nonlinear algebra.
In particular, since $\mathfrak{h}_6^{(\alpha)}$ is associative, the Jacobi identity is satisfied:
\begin{equation*}
[X,[Y,Z]] + [Y,[Z,X]] + [Z,[X,Y]] = 0 \, , \qquad \forall \,\, X,Y,Z \in \mathfrak{h}_6^{(\alpha)} \, .
\end{equation*}
In absence of the reflection operator $R$, in the limit $\alpha \to 0$, the algebra $\mathfrak{h}_6^{(\alpha)}$, defined by the commutation relations (\ref{Dunkl Two-Photon}) and (\ref{addrel}),  reduces to the standard two-photon Lie algebra $\mathfrak{h}_6$ with  Lie brackets (\ref{Two-Photon}).
\label{rem1}
\end{rem}

To construct the Casimir invariant of the algebra $\mathfrak{h}_6^{(\alpha)}$, we make use of the package \textsf{NCAlgebra}~\cite{NCAlgebra}, implemented in \textsf{Wolfram Mathematica\textsuperscript{\textregistered}}. Since the Lie algebra $\mathfrak{h}_6$ admits a Casimir invariant $\CC_{\mathfrak{h}_6} $ \eqref{eq:Casimir2}, which is a cubic polynomial in the generators, we consider all possible cubic terms obtained via complete symmetrization, defined by
\begin{equation}
	\{X,Y,Z\}:=\frac{1}{3!}\bigl(XYZ+YXZ+XZY+ZXY+YZX+ZYX\bigr)\, ,
	\qquad \forall\, X,Y,Z\in\mathfrak{h}_6^{(\alpha)} \, .
\nonumber
\end{equation}
This procedure yields a list $\ell$ consisting of $120$ elements, each of which is a fully symmetrized polynomial of degree three in the generators. Accordingly, we consider the most general cubic polynomial of the form
\begin{equation}
	P=\sum_{i=1}^{120} c_i\,\ell_i\, ,
\nonumber
\end{equation}
where the $c_i$'s are undetermined coefficients and $\ell_i$ denotes the $i$-th element of the list $\ell$. Requiring that $P$ commutes with the generating set (\ref{generators}) leads to a system of algebraic equations for the coefficients $c_i$. Solving this system and discarding trivial solutions, we obtain a nontrivial cubic symmetrized polynomial which, after reordering, turns out to be:
\begin{align}
\CC_{\mathfrak{h}_6^{(\alpha)}} &= B_+ B_- M
- A_-^2 B_+ - A_+^2 B_- - N^2 M - (1+\alpha) N M^2 + 2 N A_+ A_- \nonumber\\
&\qquad + (1+\alpha) A_+ A_- M + 3 N M
+ \alpha\!\left( N M R+\frac12 M^2 R+\frac12 M R\right)
+ \frac{\alpha^2}{2} M^2 R \, .
\label{eq:Casimir2P}
\end{align}
It is straightforward to verify that $\big[\CC_{\mathfrak{h}_6^{(\alpha)}},X\big]=0$ for all $X\in\mathfrak{h}_6^{(\alpha)}$.  In the limit $\alpha\to 0$, and in the absence of the reflection generator $R$, the above expression reduces to  the Casimir invariant $\CC_{\mathfrak{h}_6}$ given in  (\ref{eq:Casimir2}).


\subsection{$\mathfrak{h}_6^{(\alpha)}$-subalgebras  and corresponding Casimir invariants}
\label{subsec3.2}

As already mentioned, the two-photon Lie algebra $\mathfrak{h}_6$ contains several Lie subalgebras   of mathematical and physical interest~\cite{BBH09} as, for instance, those presented in (\ref{embed}). For our purposes, let us focus on six $\mathfrak{h}_6$-subalgebras which can be extended to the Dunkl framework by considering  the reflection operator $R$, all of them being 
five-dimensional  Dunkl subalgebras within the seven-dimensional  Dunkl two-photon algebra $\mathfrak{h}_6^{(\alpha)}$. These are listed in Table~\ref{table1} together  with  their non-vanishing  commutation relations and corresponding quadratic Casimir invariants (besides the common central generator $M$).

  Let us now comment and present some   properties provided by the results given in Table~\ref{table1}. Firstly, if we denote the Dunkl Heisenberg--Weyl algebra by  
  \be
  \mathfrak{h}_3^{(\alpha)}={\rm span}\{ A_+,A_-,M,R  \}  
  \label{h3a}
  \ee
  with defining relations given by
 \be
  [A_-, A_+] = M (\mathrm{Id} + \alpha R) \, , \qquad R A_\pm=-A_\pm R \, , \qquad  [M, \cdot\, ]  = 0  \, ,
 \label{h3D}
 \ee
 we find the following subalgebra embeddings:  
\begin{equation}
\begin{aligned}
	\mathfrak{h}_3^{(\alpha)}&\subset \mathfrak{h}_4^{(\alpha)} \subset \mathfrak{h}_6^{(\alpha)} \, ,\quad &
	\mathfrak{h}_3^{(\alpha)}&\subset \bar{\mathcal{G}}_\pm^{(\alpha)} \subset \mathfrak{h}_6^{(\alpha)}\, ,\quad   & \mathfrak{h}_3^{(\alpha)}&\subset \bar{\mathcal{E}}^{(\alpha)}\subset \mathfrak{h}_6^{(\alpha)}\, ,\\
	\mathfrak{h}_3^{(\alpha)}&\subset \bar{\mathcal{P}}^{(\alpha)} \subset \mathfrak{h}_6^{(\alpha)}\, ,\quad & &\mathfrak{gl}(2)^{(\alpha)}\subset \mathfrak{h}_6^{(\alpha)}\, .& &
	\label{embeddings}
	\end{aligned}
\end{equation}

Secondly, the terminology for   both Dunkl extended {\em Galilei} subalgebras   $\bar{\mathcal{G}}_\pm^{(\alpha)}$ comes from their   kinematical interpretation as the algebra of isometries  of the    $(1+1)$-dimensional Newtonian spacetime, such that for   $\bar{\mathcal{G}}_+^{(\alpha)}$, the generator $A_-$ can be regarded as the Galilean boost (resp.~$A_+$ for $\bar{\mathcal{G}}_-^{(\alpha)}$), $B_+$   as the generator of time translations (resp.~$B_-$ for $\bar{\mathcal{G}}_-^{(\alpha)}$), $A_+$   as the generator of spatial translations (resp.~$A_-$ for $\bar{\mathcal{G}}_-^{(\alpha)}$), and $M$ is a nontrivial central extension (the mass of a particle in a free kinematics), which in the Dunkl setting encompasses the reflection operator.  In this respect, for  the Dunkl extended {\em Euclidean} subalgebra $ \bar{\mathcal{E}}^{(\alpha)}$, the generator $\LL_+$ behaves as a rotation of the 2-vector $(A_+,A_-)$ on the two-dimensional Euclidean space, while for the Dunkl extended {\em Poincar\'e} subalgebra $ \bar{\mathcal{P}}^{(\alpha)}$, the generator $\LL_-$ can be seen as a Lorentzian boost acting on $(A_+,A_-)$ on the  $(1+1)$-dimensional Minkowskian spacetime. In fact, note that the generators $ \LL_+$ and $\LL_-$ defined for $\bar{\mathcal{E}}^{(\alpha)}$ and $\bar{\mathcal{P}}^{(\alpha)}$, respectively, along with the generators $N$, $M$ and $R$, span a subalgebra isomorphic to the Dunkl subalgebra  $\mathfrak{gl}(2)^{(\alpha)}$:
\begin{equation}
	[N,\LL_+]=-2\LL_-\, ,\qquad [N,\LL_-]=-2\LL_+\,  , \qquad   [\LL_-,\LL_+]= 2N+(1+\alpha)M\, .
\nonumber
\end{equation}


\begin{table}[t!]
{\small
\caption{{\small Six relevant five-dimensional subalgebras of the Dunkl two-photon algebra $\mathfrak{h}_6^{(\alpha)}$ with defining relations given by (\ref{Dunkl Two-Photon}) and (\ref{addrel}). For each case, the non-vanishing commutation rules and the (quadratic) Casimir polynomial are given.}} 
 
\begin{center} 
\begin{tabular}{l}
\hline

\hline
\\[-4pt]
$\bullet$ Dunkl harmonic oscillator subalgebra: $\mathfrak{h}_4^{(\alpha)}={\rm span}\{ N,A_+,A_-,M,R  \}$ \\[4pt]
\phantom{$\bullet$}   $[N, A_+] = A_+\qquad [N, A_-] = -A_-\qquad  [A_-, A_+] = M (\mathrm{Id} + \alpha R)\qquad R A_\pm=-A_\pm R $\\[4pt]
\phantom{$\bullet$}   $\CC_{\mathfrak{h}_4^{(\alpha)}} =A_+ A_- -NM  +\frac{\alpha}{2}MR$\\[8pt]
				
$\bullet$ Dunkl centrally extended Galilei subalgebra: $\bar{\mathcal{G}}_+^{(\alpha)}={\rm span}\{  A_+,A_-,B_+,M,R  \}$ \\[4pt]
\phantom{$\bullet$}   $[A_-, B_+]  = 2A_+ \qquad \quad\! [A_-, A_+] = M (\mathrm{Id} + \alpha R)\qquad R A_\pm=-A_\pm R $\\[4pt]
\phantom{$\bullet$}   $\CC_{\bar{\mathcal{G}}_+^{(\alpha)} } =A_+^2-B_+M  $\\[8pt]
				
$\bullet$ Dunkl centrally extended Galilei subalgebra: $\bar{\mathcal{G}}_-^{(\alpha)}={\rm span}\{  A_+,A_-,B_-,M,R  \}$ \\[4pt]
\phantom{$\bullet$}   $[A_+, B_-]  = -2A_-  \qquad  [A_-, A_+] = M (\mathrm{Id} + \alpha R)\qquad R A_\pm=-A_\pm R $\\[4pt]
\phantom{$\bullet$}   $\CC_{\bar{\mathcal{G}}_-^{(\alpha)} } =A_-^2-B_-M   $\\[8pt]
				
$\bullet$ Dunkl centrally extended Euclidean subalgebra: $\bar{\mathcal{E}}^{(\alpha)}={\rm span}\big\{ \LL_+:=\frac 12(B_-+B_+), A_+,A_-, M,R \big \}$ \\[4pt]
\phantom{$\bullet$}   $[\LL_+, A_+]  =   A_-  \qquad[\LL_+, A_-]  = - A_+    \qquad [A_-, A_+] = M (\mathrm{Id} + \alpha R)\qquad R A_\pm=-A_\pm R $\\[4pt]
\phantom{$\bullet$}   $\CC_{ \bar{\mathcal{E}}^{(\alpha)} } =  A_-^2+A_+^2-2  \LL_+ M $\\[8pt]
				
$\bullet$ Dunkl centrally extended Poincar\'e subalgebra: $\bar{\mathcal{P}}^{(\alpha)}={\rm span}\big\{ \LL_-:=\frac 12(B_- -B_+), A_+,A_-, M,R  \big\}$ \\[4pt]
\phantom{$\bullet$}   $[ \LL_-, A_+]  =   A_-   \qquad [ \LL_-, A_-]  =   A_+  \qquad\quad\! [A_-, A_+] = M (\mathrm{Id} + \alpha R)\qquad R A_\pm=-A_\pm R $\\[4pt]
\phantom{$\bullet$}   $\CC_{ \bar{\mathcal{P}}^{(\alpha)} } =  A_-^2-A_+^2-  2   \LL_- M$\\[8pt]
				
$\bullet$ Dunkl  $\mathfrak{gl}(2) $ subalgebra: $\mathfrak{gl}(2)^{(\alpha)}={\rm span}\{ N,  B_+ ,  B_-, M,R  \}$ \\[4pt]
\phantom{$\bullet$}   $[N, B_+]  = 2B_+ \qquad [N, B_-]  = -2B_-\qquad [B_-, B_+]  = 4N+2(1+\alpha)M $\\[4pt]
\phantom{$\bullet$}   $\CC_{\mathfrak{gl}(2)^{(\alpha)} } =   B_+ B_- -N^2 + 2 N -(1+\alpha) NM   $\\[8pt]
				
\hline
				
\hline
\end{tabular}
\end{center} 
\label{table1}
}
\end{table}


Thirdly, if we   compute the   commutators corresponding to the $\mathfrak{h}_6^{(\alpha)}$-generators outside of  each of the  first  five  subalgebras displayed in Table~\ref{table1}  with their Casimir invariant, we find that the results involve other Casimir polynomials in this set, explicitly
\begin{equation}
	\begin{aligned}
		\mathfrak{h}_4^{(\alpha)}&\!: \ \ & \big[\CC_{\mathfrak{h}_4^{(\alpha)}} ,B_+\big] &= 2 \CC_{ \bar{\mathcal{G}}_+^{(\alpha)} } \, , \quad &  \big[\CC_{\mathfrak{h}_4^{(\alpha)}} ,B_-\big] &=-2 \CC_{ \bar{\mathcal{G}}_-^{(\alpha)} } \, .\\[1pt]
		\bar{\mathcal{G}}_+^{(\alpha)}&\!: \quad    & \big[\CC_{ \bar{\mathcal{G}}_+^{(\alpha)}  } ,N\big] &=- 2\CC_{ \bar{\mathcal{G}}_+^{(\alpha)} } \, , \quad &  \big[\CC_{ \bar{\mathcal{G}}_+^{(\alpha)} } ,B_-\big] &= -4 \CC_{\mathfrak{h}_4^{(\alpha)}}  - 2 M +2(1+\alpha) M^2 \, .\\[1pt]
		\bar{\mathcal{G}}_-^{(\alpha)} &\!: \quad    & \big[\CC_{ \bar{\mathcal{G}}_-^{(\alpha)}  } ,N\big] &=  2\CC_{ \bar{\mathcal{G}}_-^{(\alpha)}}  \, , \quad &  \big[\CC_{ \bar{\mathcal{G}}_-^{(\alpha)} } ,B_+\big] &= 4 \CC_{\mathfrak{h}_4^{(\alpha)}}  + 2 M -2(1+\alpha) M^2 \, .\\[1pt]
		\bar{\mathcal{E}}^{(\alpha)}&\!: \quad    & \big[\CC_{ \bar{\mathcal{E}}^{(\alpha)} } ,N\big] &=  2 \CC_{ \bar{\mathcal{P}}^{(\alpha)} } \, , \quad &  \big[\CC_{ \bar{\mathcal{E}}^{(\alpha)} } ,\LL_-\big] &=  -4 \CC_{\mathfrak{h}_4^{(\alpha)}}  - 2 M +2(1+\alpha) M^2 \, .\\[1pt]
		\bar{\mathcal{P}}^{(\alpha)}&\!: \quad    & \big[\CC_{ \bar{\mathcal{P}}^{(\alpha)} } ,N\big] &=  2 \CC_{ \bar{\mathcal{E}}^{(\alpha)} } \, , \quad &  \big[\CC_{ \bar{\mathcal{P}}^{(\alpha)} } ,\LL_+\big] &=   4 \CC_{\mathfrak{h}_4^{(\alpha)}}  + 2 M -2(1+\alpha) M^2 \, . 
		\label{comcash6}
	\end{aligned}
\end{equation}
By contrast, for ${\mathfrak{gl}(2)^{(\alpha)} } $ there is no kind of such relations as we find that
\begin{equation}
	\begin{aligned}
		\big[\CC_{\mathfrak{gl}(2)^{(\alpha)} } ,A_+\big] &=-2 N A_+ -A_+ + 2 A_- B_+ - (1+\alpha) A_+ M \, , \\[4pt]
		\big[\CC_{\mathfrak{gl}(2)^{(\alpha)} } ,A_-\big] &= \phantom{+}2 N A_- -A_- -2 A_+ B_- + (1+\alpha) A_- M  \, . 
	\end{aligned}
	\label{comcash6b}
\end{equation}


Finally, the Casimir invariants $\CC_{\bar{\mathcal{G}}_\pm^{(\alpha)}}$, $\CC_{\mathfrak{h}_4^{(\alpha)}}$ and $M$ allow us to obtain an alternative expression involving the cubic Casimir $\CC_{\mathfrak{h}_6^{(\alpha)}} $   (\ref{eq:Casimir2P}), as the quartic polynomial given by
\begin{align}
		\frac{1}{2}\left\{\CC_{\bar{\mathcal{G}}_+^{(\alpha)}} ,\CC_{\bar{\mathcal{G}}_-^{(\alpha)}}\right\}  -\left(\CC_{\mathfrak{h}_4^{(\alpha)}}\right)^2-\CC_{\mathfrak{h}_4^{(\alpha)}} M-\left(1-\frac{\alpha^2}{4}\right)M^2&+(1+\alpha)\CC_{\mathfrak{h}_4^{(\alpha)}}M^2 -(1+\alpha)M^3=\CC_{\mathfrak{h}_6^{(\alpha)}}  M\, , 
\label{quarticcash6}
\end{align}
where $\{\cdot, \cdot\}$ denotes the anticommutator. From this relation, we observe that if we apply the limit  $\alpha \to 0$ and consider the Lie--Poisson counterpart of $\mathfrak{h}_6^{(\alpha)}$, {\em  i.e.}~with Lie--Poisson brackets given formally by (\ref{Dunkl Two-Photon}) with $\alpha=0$, discarding lower order terms, we recover the expression presented in~\cite{BBH09}, namely the quartic relation:
	\begin{equation}
		\CC_{\bar{\mathcal{G}}_+^{(0)}}  \CC_{\bar{\mathcal{G}}_-^{(0)}}  -\left(\CC_{\mathfrak{h}_4^{(0)}}\right)^2 = \CC_{\mathfrak{h}_6^{(0)}}  M \, .
\nonumber
	\end{equation}
As expected,   the relation (\ref{quarticcash6}) is the symmetrized version, with one additional reflection, of the $\mathfrak{h}_6$-Poisson Casimir belonging to the symmetric algebra $S(\mathfrak{h}_6)$ of $\mathfrak{h}_6$ (\ref{Two-Photon}).


\subsection{Dunkl--Fock space representation and  Dunkl harmonic oscillator}
\label{subsec3.3}

A realization of $\mathfrak{h}_6^{(\alpha)}$ on the number-state Hilbert space of the Dunkl oscillator, spanned by the orthonormal states $\{\ket{n}_\mu\}_{n=0}^{\infty}$, is obtained in terms of the bosonic operators $\hat{a}$, $\hat{a}^\dagger$ and the reflection operator $\hat{R}$ in the form
\begin{equation}
\begin{aligned}
	\hat{N}&=\hat{a}^\dagger \hat{a}-\mu\left(\hat{\mathds{1}}-(-1)^{\hat{N}}\right) \, ,\quad &\hat{A}_+&=\hat{a}^\dagger \, , \quad &\hat{A}_-&=\hat{a}\, ,\\
	\hat{M}&=\hat{\mathds{1}}\, ,\qquad\ \hat{B}_+=\big(\hat{a}^\dagger\big)^2 \, , \quad& \hat{B}_-&=\hat{a}^2 \,, \quad& \hat{R}&=(-1)^{\hat{N}}\, ,
\end{aligned}
 \label{eq:reps1}
 \end{equation}
which satisfy the relations \eqref{Dunkl Two-Photon} and \eqref{addrel}  provided that $\alpha=2\mu>-1$.  Hereafter, the notation $(-1)^{\hat{N}}\equiv \exp\big(\imath \pi \hat{N} \big)$ will be assumed. The explicit action on the states is given by
\begin{equation}
	\hat{N} \ket{n}_\mu=n \ket{n}_\mu \, ,\qquad \hat{M} \ket{n}_\mu=\ket{n}_\mu \, ,\qquad  \hat{R} \ket{n}_\mu=(-1)^n \ket{n}_\mu \, ,
	\label{eq:actionNMR}
\end{equation}
together with
\begin{equation}
\begin{aligned}
	\hat{A}_+ \ket{n}_\mu&=\sqrt{[n+1]_\mu}\, \ket{n+1}_\mu\, ,\quad 	&\hat{A}_- \ket{n}_\mu&=\sqrt{[n]_\mu} \,  \ket{n-1}_\mu \, ,  \\[2pt]
	\hat{B}_+ \ket{n}_\mu&=\sqrt{[n+1]_\mu [n+2]_\mu}\, \ket{n+2}_\mu \, ,\quad 	&\hat{B}_- \ket{n}_\mu&=\sqrt{[n]_\mu [n-1]_\mu}\, \ket{n-2}_\mu  \, ,
	\label{eq:actionB}
	\end{aligned}
\end{equation}
where we have introduced the $\mu$-numbers~\cite{CH} defined by
\begin{equation}
	[n]_\mu:=n+\mu\big(1-(-1)^n \big) \, .
	\label{eq:munumbers}
\end{equation} 
Notice that, with this notation, we have that
\begin{equation}
	\hat{a}^\dagger \hat{a}=\big[\hat{N}\big]_{\mu} \, , \qquad \hat{a} \hat{a}^\dagger=\big[\hat{N}+\hat{\mathds{1}}\big]_{\mu} \, ,
\nonumber
\end{equation}
which is consistent with the following commutation relation:
\begin{equation}
	\big[\hat{a}, \hat{a}^\dagger\big]=\hat{a} \hat{a}^\dagger-\hat{a}^\dagger \hat{a}=\big[\hat{N}+\hat{\mathds{1}}\big]_{\mu}-\big[\hat{N}\big]_{\mu}=\hat{\mathds{1}}+2 \mu (-1)^{\hat{N}}= \hat{\mathds{1}}+2 \mu \hat{R}\, .
	\label{eq:commrel}
\end{equation}

 Consequently, the realization   (\ref{eq:reps1}), the action on the  Dunkl--Fock space determined by  (\ref{eq:actionNMR})--(\ref{eq:munumbers}) and the Dunkl Heisenberg--Weyl commutator (\ref{eq:commrel}) turn out to be the extensions to the Dunkl setting, with a reflection, of the standard expressions (\ref{eq:reps0}), (\ref{eq:action0}) and (\ref{h3}), respectively, which are recovered in the limit $\mu \to 0$, as expected. In this regard, it should be noted that the Dunkl Heisenberg--Weyl subalgebra $\mathfrak{h}_3^{(2\mu)}$ (\ref{h3a})  requires the presence of the reflection operator $\hat{R}$, as shown in (\ref{eq:commrel}),  so that the realization (\ref{eq:reps1}) yields the one corresponding   to $\mathfrak{h}_3^{(2\mu)}$  with the defining relations (\ref{h3D}).

In addition, the action of the reflection operator on the states $\ket{n}_{\mu}$, written in (\ref{eq:actionNMR}), can be understood as
\begin{align}
	\hat{R} \ket{n}_{\mu}=(-1)^{\hat{N}}\ket{n}_{\mu}=\exp\big(\imath \pi \hat{N} \big)\ket{n}_{\mu}&=\sum_{k=0}^\infty\frac{\big(\imath \pi \hat{N}\big)^k}{k!}\ket{n}_{\mu}=\sum_{k=0}^\infty\frac{(\imath \pi)^k}{k!}\hat{N}^k\ket{n}_{\mu} \nonumber\\[2pt]
	&=\sum_{k=0}^\infty\frac{(\imath \pi)^k}{k!}n^k\ket{n}_{\mu}  
	 =\sum_{k=0}^\infty\frac{(\imath \pi n)^k}{k!}\ket{n}_{\mu}\nonumber \\[2pt]
	&=\exp(\imath \pi n)\ket{n}_{\mu}  =(-1)^n\ket{n}_{\mu} \, ,
\nonumber
\end{align}
where $\exp(\imath \pi n)=\cos (\pi n)+\imath \sin (\pi n)$. The value of the cubic Casimir \eqref{eq:Casimir2P} in the given Dunkl representation yields
\begin{equation}
	\hat{\CC}_{\mathfrak{h}_6^{(2 \mu)}}   \ket{n}_\mu=2\big(\mu(\mu-1)-1\big)\ket{n}_\mu \, .
	\label{eq:Cas}
\end{equation}
Remember that $\alpha = 2\mu>-1$.
Let us also notice that, under such a representation, the only nonzero values of the quadratic Casimirs of the $\mathfrak{h}_6^{(\alpha)}$-subalgebras listed in Table~\ref{table1}   are
\begin{align}
\hat{\CC}_{\mathfrak{h}_4^{(2 \mu)}}\ket{n}_{\mu}=\mu \ket{n}_{\mu} \, , \qquad \hat{\CC}_{\mathfrak{gl}(2)^{(2\mu)}}\ket{n}_{\mu}=\mu\big((-1)^n-1\big)\ket{n}_{\mu} \, .
\label{eq:subCas}
\end{align}

A complementary realization of the representation \eqref{eq:reps1} can be obtained directly  in the coordinate space. Let $\{\langle x|\}_{x\in\mathbb{R}}$ denotes the generalized eigenbasis of the position operator $\hat{x}$, so that states in the number basis admit the coordinate-space wavefunctions:
\begin{equation}
	\psi_n(x) := \braket{x|n} \, .
\nonumber
\end{equation}
In this representation, the annihilation and creation operators can be expressed in terms of the Dunkl derivative $\mathcal{D}_x$~\cite{Dunkl1989,Ros2003, Dunkl2012} and the coordinate multiplication operator. Recall that the one-dimensional Dunkl derivative associated with the reflection operator, $\hat{R} f(x)=f(-x)$, is defined by
\begin{equation}
	\mathcal{D}_x f(x) := \partial_x f(x)+ \frac{\mu}{x}\big(f(x)-f(-x)\big)\, , \qquad \partial_x:=\frac{\text{d}}{\text{d}x} \, .
	\label{Dderivative}
\end{equation}
Under this realization, the basic ladder operators read
\begin{equation}
	\hat{a}=\frac{1}{\sqrt{2}}\big(x+\mathcal{D}_x\big)\, ,\qquad 
	\hat{a}^\dagger=\frac{1}{\sqrt{2}}\big(x-\mathcal{D}_x\big) \, .
	\label{ladder}
\end{equation}
The wavefunctions $\psi_n(x)$ are the orthonormal eigenfunctions of the Dunkl harmonic oscillator Hamiltonian in coordinate space and can be written explicitly in terms of the generalized Hermite polynomials $H_n^{(\mu)}(x)$ as \cite{MSSM, Rosenblum, GIVZ2013I}
\begin{equation}
	\psi_n^{(\mu)}(x) :=\braket{x|n}_\mu= \mathrm{e}^{-\frac{x^2}{2}}\, H_n^{(\mu)}(x) \, ,
\nonumber
\end{equation}
where we have introduced the generalized Hermite polynomials defined by
\begin{equation}
	H_{2k+p}^{(\mu)}(x):=(-1)^k \sqrt{\frac{k!}{\Gamma(k+p+\mu+1/2)}} \, x^p L_k^{(\mu+p-1/2)}(x^2) \, , \qquad p=0,1 \, .
\nonumber
\end{equation}
Here, $L_k^{(a)}(x)$ with $a>-1$ and $k=0,1,2,\dots$, are the Laguerre polynomials. Thus, the eigenfunctions are defined in the even and odd sectors as follows:
\begin{equation}
	\psi_n^{(\mu)}(x)=
	\left\{
	\begin{aligned}
	 &(-1)^{k}\sqrt{\dfrac{k!}{\Gamma\left(k+\mu+\frac{1}{2}\right)}}\, \mathrm{e}^{-\frac{x^2}{2}}L_{k}^{(\mu-1/2)}(x^2) \, , \hskip 0.9cm n=2k\, ,\\[6pt]
	&(-1)^{k}\sqrt{\dfrac{k!}{\Gamma\left(k+\mu+\frac{3}{2}\right)}}\, x \, \mathrm{e}^{-\frac{x^2}{2}}L_{k}^{(\mu+1/2)}(x^2) \, , \hskip 0.95 cm n=2k+1 \, . 
	\end{aligned}\right. 
\nonumber
\end{equation}
These eigenfunctions are orthonormal in the Hilbert space $L^2\big(\mathbb{R}, |x|^{2\mu}\text{d}x \big)$ with respect to the inner product defined by  (see, e.g., \cite{Quesne2024}):
\begin{equation}
	\langle{f,g}\rangle_{\mu}:=\int_{\mathbb{R}} f(x)^* g(x) |x|^{2\mu} \text{d}x \, ,
	\label{innerP}
\end{equation}
which means that the following equality holds
\begin{equation}
	\big\langle{\psi_n^{(\mu)},\psi_m^{(\mu)}}\big\rangle_{\mu}=\delta_{n,m} \, ,
	\label{eq:orthonormality}
\end{equation}
which can be verified directly. Then, the ladder relations can be easily checked
\begin{equation}
	\hat{a}\,\psi_n^{(\mu)}(x)=\sqrt{[n]_{\mu}}\,\psi^{(\mu)}_{n-1}(x)\, ,\qquad
	\hat{a}^\dagger \psi_n^{(\mu)}(x)=\sqrt{[n+1]_\mu}\,\psi^{(\mu)}_{n+1}(x)\, ,
	\nonumber
\end{equation}
which mirror exactly the algebraic action derived in the number-state representation. Let us remark that in this framework, the Hamiltonian of the Dunkl harmonic oscillator is given by 
\begin{equation}
	\hat{H}=-\frac{\mathcal{D}_x^2}{2}+\frac{x^2}{2}=\frac{1}{2}\big\{\hat{a}^\dagger, \hat{a}\big\}=\hat{N}+\left(\mu+\frac{1}{2}\right)\hat{\mathds{1}}
	\label{eq:hosc}
\end{equation}
(in units such as $m=\omega= \hbar=1$) and, therefore,  its spectrum is found to be
\begin{equation}
	\hat{H} \psi_n^{(\mu)}(x)=\left(n+\mu+\frac{1}{2}\right)\psi_n^{(\mu)}(x) \, .
	\label{eq:Ham}
\end{equation}

The algebra $\mathfrak{h}_6^{(\alpha)}$, for $\alpha=2\mu>-1$, then finds  is complete realization in the coordinate space as
\begin{equation}
	\hat{N}\, \psi_{n}^{(\mu)}(x)=n \,\psi^{(\mu)}_{n}(x) \, ,\quad\ \hat{M} \psi_{n}^{(\mu)}(x)=\psi_{n}^{(\mu)}(x) \, ,\quad\  \hat{R} \,\psi_{n}^{(\mu)}(x)=(-1)^n \psi_{n}^{(\mu)}(x)\, ,
\nonumber
\end{equation}
along with 
\begin{equation}
\begin{aligned}
	\hat{A}_+\psi_{n}^{(\mu)}(x)&=\sqrt{[n+1]_\mu} \, \psi_{n+1}^{(\mu)}(x) \, ,\quad &	\hat{A}_- \psi_{n}^{(\mu)}(x)&=\sqrt{[n]_\mu}\,\psi^{(\mu)}_{n-1}(x) \, ,\\[2pt]
	\hat{B}_+ \psi_{n}^{(\mu)}(x)&=\sqrt{[n+1]_\mu [n+2]_\mu}\, \psi^{(\mu)}_{n+2}(x)\, ,\quad &	\hat{B}_- \psi_{n}^{(\mu)}(x)&=\sqrt{[n]_\mu [n-1]_\mu}\,\psi^{(\mu)}_{n-2}(x) \, ,
\nonumber
\end{aligned}	
\end{equation}
where the differential-difference realization of the operators can be obtained straightforwardly from  the initial Fock realization of $\mathfrak{h}_6^{(\alpha)}$ \eqref{eq:reps1}, taking into account \eqref{ladder} and \eqref{eq:hosc} as well. Obviously, within this framework, the equations \eqref{eq:Cas} and \eqref{eq:subCas} are simply reformulated as
\begin{equation}
	\hat{\CC}_{\mathfrak{h}_6^{(2 \mu)}}   \psi_n^{(\mu)}(x)=2\big(\mu(\mu-1)-1 \big)  \psi_n^{(\mu)}(x)\, ,
\nonumber
\end{equation}
\begin{equation}
	 \hat{\CC}_{\mathfrak{h}_4^{(2 \mu)}}  \psi_n^{(\mu)}(x)=\mu \, \psi_n^{(\mu)}(x) \, , \qquad \hat{\CC}_{\mathfrak{gl}(2)^{(2\mu)}}  \psi_n^{(\mu)}(x)=\mu \big((-1)^n-1\big) \psi_n^{(\mu)}(x) \, .
\nonumber
\end{equation}

We conclude by observing that when $\mu = 0$, in the absence of the reflection operator $\hat{R} $,  the entire construction above  collapses to the standard one, which is already well-known for the two-photon Lie algebra $\mathfrak{h}_6$~\cite{ZFG, Brif}. In this case, in fact, we find that
\begin{equation}
	\psi_n^{(\mu)}(x)\big|_{\mu=0}\equiv\psi_n(x)=\frac{1}{\pi^{1/4}\sqrt{2^n n!} }\, \mathrm{e}^{-\frac{x^2}{2}}H_n(x) \, ,  
	\label{eq:collapse}
\end{equation}
where $H_n(x)$ are the Hermite polynomials, as it can be directly checked. These eigenfunctions  are orthonormal in the usual Hilbert space $L^2(\mathbb{R}, \text{d}x)$ with respect to the  inner product (see (\ref{innerP})):
\begin{equation}
	\langle{f,g}\rangle:=\int_{\mathbb{R}} f(x)^* g(x) \, \text{d}x \, ,
	\nonumber
\end{equation}
and the following equality holds
\begin{equation}
	\langle \psi_n,\psi_m\rangle=\delta_{n,m} \, ,
\nonumber
\end{equation}
which is formally the same as in the reflection  $\mu$-case (\ref{eq:orthonormality}). Thus,   the eigenfunctions  (\ref{eq:collapse}) are those associated with the spectral problem of the standard quantum harmonic oscillator. 
Analogously, for the states  $\ket{n}_\mu|_{\mu=0}\equiv \ket{n}$, we have the orthonormality condition $\braket{n|m}=\delta_{n,m}$, and the  Dunkl--Fock space reduces  to the one corresponding to the standard quantum harmonic oscillator.


\subsection{Dunkl--Fock--Bargmann (holomorphic) representation}
\label{subsec3.4}

Another representation that is commonly used in connection with the two-photon Lie algebra $\mathfrak{h}_6$~\cite{ZFG,Brif} is the so-called Fock--Bargmann representation~\cite{Bargmann}. In the present Dunkl framework, the generators of $\mathfrak{h}_6^{(\alpha)}$ (\ref{generators}) act on the space of entire analytic functions $f(z)$ as differential-difference operators, according to the following holomorphic realization
\begin{equation}
\begin{aligned}
\bar{N} &= z\,\mathcal{D}_z - \mu\big(\bar{\mathds{1}} - \exp(\imath \pi z \partial_z)\big)\, , 
\qquad  \bar{A}_+ = z\,  ,  \qquad  \bar{A}_- = \mathcal{D}_z\,  , \\[2pt]
\bar{M} &= \bar{\mathds{1}}\,  ,  \qquad  \bar{B}_+ = z^2\,  ,  \qquad  \bar{B}_- = \mathcal{D}_z^{\,2}\,  , 
\qquad  \bar{R} = \exp(\imath \pi z \partial_z)\,  ,
\end{aligned}
\label{eq:reps2z}
\end{equation}
where $\bar{\mathds{1}}$ denotes the identity operator on the space of entire functions and $\mathcal{D}_z$ is the Dunkl derivative defined in (\ref{Dderivative}), thereby generalizing the standard representation (\ref{eq:reps2z0}).
It can be shown by direct calculation that  \eqref{eq:reps2z} satisfies the defining relations \eqref{Dunkl Two-Photon} and \eqref{addrel},   provided that $\alpha=2\mu>-1$. Let us notice that, after expanding the Dunkl derivative, the number operator simplifies to $\bar{N}=z\partial_z$, as the additional terms cancel.

To transform the holomorphic realization into a Hilbert space representation, we endow the space of entire functions with a suitable scalar product. We denote by $\mathrm{d}^2 z$ the Lebesgue measure on $\mathbb{C}\cong\mathbb{R}^2$, namely
\begin{equation}
\mathrm{d}^2 z := \mathrm{d}x\,\mathrm{d}y \qquad \text{for}\quad z=x+\imath y \, ,
\nonumber
\end{equation}
which in polar coordinates $z = r \mathrm{e}^{\imath\theta}$ reads as
\begin{equation}
\mathrm{d}^2 z = r\,\mathrm{d}r\,\mathrm{d}\theta \, .
\nonumber
\end{equation}
Then, we introduce the radial densities \cite{Sontz1, Sontz2} (see, also, ref.~\cite{Ghaz2022}):
\begin{equation}
\begin{aligned}
\nu_{e,\mu}(r) &:= c_\mu\, r^{2\mu+1} K_{\mu-\frac12}(r^2)\,  ,
\qquad c_\mu := \frac{2^{\frac12-\mu}}{\pi\,\Gamma\!\left(\mu+\tfrac12\right)}\, ,
\\
\nu_{o,\mu}(r) &:= c_\mu\, r^{2\mu+1} K_{\mu+\frac12}(r^2)\, , \quad 
\nonumber
\end{aligned}
\end{equation}
where $K_\alpha$ is the modified Bessel function of   second kind. These densities define two (even/odd) measures on $\mathbb{C}$:
\begin{equation}
\mathrm{d}\nu_{e,\mu}(z) := \nu_{e,\mu}(|z|)\,\mathrm{d}^2 z\,  , 
\qquad
\mathrm{d}\nu_{o,\mu}(z) := \nu_{o,\mu}(|z|)\,\mathrm{d}^2 z\, .
\nonumber
\end{equation}
And  every analytic function $f(z)$ decomposes into even and odd parts: 
\begin{equation}
f(z)=f_e(z)+f_o(z)\, , \qquad
f_e(z):=\frac{f(z)+f(-z)}{2}\, , \qquad
f_o(z):=\frac{f(z)-f(-z)}{2}\, .
\label{eq:evenodddecomp-final}
\end{equation}
We define the Dunkl--Fock--Bargmann space as
\begin{equation}
\mathcal{H}_\mu^{(B)} := \left\{ f \ \text{analytic on }\mathbb{C}\ :\ 
\|f\|_{\mu,B}^2 < \infty \right\},
\label{Hspace}
\end{equation}
where 
\begin{equation}
\|f\|_{\mu,B}^2 :=
\int_{\mathbb{C}} |f_e(z)|^2\,\mathrm{d}\nu_{e,\mu}(z) +
\int_{\mathbb{C}} |f_o(z)|^2\,\mathrm{d}\nu_{o,\mu}(z)\, , \qquad
f=f_e+f_o\, ,
\label{normFB}
\end{equation}
 with scalar product defined by
\begin{equation}
\langle f,g\rangle_{\mu,B} :=
\int_{\mathbb{C}} f_e(z)^*\,g_e(z)\,\mathrm{d}\nu_{e,\mu}(z) +
\int_{\mathbb{C}} f_o(z)^*\,g_o(z)\,\mathrm{d}\nu_{o,\mu}(z)\, .
\nonumber
\end{equation}
In other words, the space $\mathcal{H}^{(B)}_\mu$ (\ref{Hspace}) can be viewed as the subspace of entire functions of
\begin{equation}
L^2(\mathbb{C},\mathrm{d}\nu_{e,\mu}) \oplus L^2(\mathbb{C},\mathrm{d}\nu_{o,\mu}) \, ,
\nonumber
\end{equation}
with the reflection operator $\bar{R}$ acting as $\bar{R} f(z) = f(-z)$ and leaving the even and odd sectors invariant.

  In analogy with the standard Fock--Bargmann representation, we introduce the monomials:
\begin{equation}
\phi_n^{(\mu)}(z) := \frac{z^n}{\sqrt{[n]_\mu!}} \, , 
\qquad n=0,1,2,\dots,
\label{eq:monomials-final}
\end{equation}
where the  $\mu$-number $[n]_\mu$ is given in \eqref{eq:munumbers}, whereas the $\mu$-factorial is  defined by~\cite{TVZ}
$$
[n]_\mu!:=[1]_\mu [2]_\mu \dots [n]_\mu
\, .
$$
The functions (\ref{eq:monomials-final}) belong to $\mathcal{H}_\mu^{(B)}$ and form a complete orthonormal set,
as one can verify through the orthonormality relation:
\begin{equation}
\big\langle \phi_n^{(\mu)},\phi_m^{(\mu)} \big\rangle_{\mu,B}
= \delta_{n,m}\, .
\label{eq:orthonormal-B-final}
\end{equation}

  In the definition of the Dunkl--Fock--Bargmann  Hilbert space the norm (\ref{normFB}) contains two integrals, one weighted by $\mathrm{d}\nu_{e,\mu}$ and the other by $\mathrm{d}\nu_{o,\mu}$. This is necessary because a generic analytic function $f$ decomposes into its even and odd parts, $f=f_e+f_o$ (\ref{eq:evenodddecomp-final}), and each component must be integrated with the corresponding measure. For monomials the situation simplifies: $z^n$ is purely even when $n$ is even and purely odd when $n$ is odd:
\[
(z^n)_e=
\begin{cases}
z^n\, ,& n \ \text{even}\, ,\\
0\, ,& n \ \text{odd}\, ,
\end{cases}
\qquad
(z^n)_o=
\begin{cases}
0\, ,& n \ \text{even}\, ,\\
z^n\, ,& n \ \text{odd}\, ,
\end{cases}
\]
and only one of the two integrals contributes. The scalar product of monomials therefore reduces to:
\[
\langle z^n , z^m \rangle_{\mu,B} =
\left\{
\begin{aligned}
 &\displaystyle \int_{\mathbb{C}} |z|^{2n}\,\mathrm{d}\nu_{e,\mu}(z)\, ,\quad & n&=m \ \text{even} \, , \\[4pt]
&\displaystyle \int_{\mathbb{C}} |z|^{2n}\,\mathrm{d}\nu_{o,\mu}(z)\, , \quad &n&=m \ \text{odd} \, , \\[4pt]
&0\, ,\quad& n&\neq m \,.
\end{aligned}\right. 
\]
In particular, choosing the normalization as in \eqref{eq:monomials-final} one obtains \eqref{eq:orthonormal-B-final}, which confirms that the normalized monomials form an orthonormal basis of the Dunkl--Fock--Bargmann space.

Observe that for $\mu=0$, one has that $\nu_{e,0}=\nu_{o,0}=\pi^{-1}\mathrm{e}^{-|z|^2}$ and the space $\mathcal{H}_\mu^{(B)}$ reduces to the standard Fock--Bargmann space of the harmonic oscillator \cite{Hall}. We recall that the usual Fock--Bargmann space associated with the standard quantum harmonic oscillator is the Hilbert space defined by
\begin{equation}
\mathcal{H}^{(B)}
:= \left\{ f \,\, \text{analytic on }\mathbb{C}\\ :\ 
\|f\|_{B}^2 < \infty \right\},
\nonumber
\end{equation}
where
\begin{equation}
\|f\|_{B}^2
:= \frac{1}{\pi}\int_{\mathbb{C}} |f(z)|^2 \mathrm{e}^{-|z|^2}\,\mathrm{d}^2 z \, .
\nonumber
\end{equation}
The corresponding inner product is then given by

\begin{equation}
\langle f,g\rangle_{B} :=
\frac{1}{\pi}\int_{\mathbb{C}} f(z)^*\,g(z)\,
\mathrm{e}^{-|z|^2}\,\mathrm{d}^2 z\, .
\nonumber
\end{equation}
In this space, the monomials
\begin{equation}
\phi_n(z) := \frac{z^n}{\sqrt{n!}} \, , \qquad n=0,1,2,\dots,
\nonumber
\end{equation}
form an orthonormal basis
\begin{equation}
\langle \phi_n,\phi_m\rangle_{B} = \delta_{n,m}\, .
\nonumber
\end{equation}

Now, based on the Dunkl--Fock--Bargmann representation (\ref{eq:reps2z}), the basic ladder operators read
\begin{equation}
\bar{a}=\mathcal{D}_z \, , \qquad
\bar{a}^\dagger=z \, ,
\nonumber
\end{equation}
and the Hamiltonian of the Dunkl harmonic oscillator results in
\begin{equation}
\bar{H}=z \mathcal{D}_z+ \mu \bar R +\frac{1}{2}\, \bar{\mathds{1}}=\frac{1}{2}\big\{\bar{a}^\dagger, \bar{a}\big\}=\bar{N}+\left(\mu+\frac{1}{2}\right)\bar{\mathds{1}} \, ,
\nonumber
\end{equation}
(to be compared with (\ref{eq:hosc})) with energy spectrum given by
\begin{equation}
\bar{H} \phi_n^{(\mu)}(z)=\left(n+\mu+\frac{1}{2}\right)\phi_n^{(\mu)}(z) \, ,
\nonumber
\end{equation}
which is consistent with (\ref{eq:Ham}).

 The algebra $\mathfrak{h}_6^{(\alpha)}$, for $\alpha=2 \mu>-1$, finds  its complete Dunkl--Fock--Bargmann representation as follows:
\begin{equation}
\bar{N}\, \phi_{n}^{(\mu)}(z)=n\, \phi^{(\mu)}_{n}(z)\, ,\qquad \bar{M} \phi_{n}^{(\mu)}(z)=\phi_{n}^{(\mu)}(z) \, ,\qquad  \bar{R} \,\phi_{n}^{(\mu)}(z)=(-1)^n \phi_{n}^{(\mu)}(z) \, ,
\nonumber
\end{equation}
together with
\begin{equation}
\begin{aligned}
\bar{A}_+\phi_{n}^{(\mu)}(z)&=\sqrt{[n+1]_\mu}\,\phi_{n+1}^{(\mu)}(z) \, ,\quad &\bar{A}_- \phi_{n}^{(\mu)}(x)&=\sqrt{[n]_\mu}\,\phi^{(\mu)}_{n-1}(x) \, , \\[2pt]
 \bar{B}_+ \phi_{n}^{(\mu)}(z)&=\sqrt{[n+1]_\mu [n+2]_\mu}\,\phi^{(\mu)}_{n+2}(z)\, ,\quad &\bar{B}_- \phi_{n}^{(\mu)}(z)&=\sqrt{[n]_\mu [n-1]_\mu}\,\phi^{(\mu)}_{n-2}(z) \, .
\nonumber
\end{aligned}
\end{equation}
The value of the cubic Casimir \eqref{eq:Casimir2P} in the above representation turns out to be
\begin{equation}
\bar{C}_{\mathfrak{h}_6^{(2 \mu)}}\phi^{(\mu)}_{n}(z) =2\big(\mu(\mu-1)-1 \big)\phi^{(\mu)}_{n}(z) \, .
\nonumber
\end{equation}
Again, the only nonzero values of the Casimirs of the subalgebras in Table~\ref{table1}  are given by
\begin{align}
\bar{\CC}_{\mathfrak{h}_4^{(2 \mu)}}\phi^{(\mu)}_{n}(z) =\mu\, \phi^{(\mu)}_{n}(z)  \, , \qquad \bar{\CC}_{\mathfrak{gl}(2)^{(2\mu)}}\phi^{(\mu)}_{n}(z) =\mu\big((-1)^n-1\big)\phi^{(\mu)}_{n}(z)  \, .
\nonumber
\end{align}

At this point, we consider it useful   to present the Dunkl counterpart of the differential equation  (\ref{eq:defeqmu01})  that produces the two-photon algebra eigenstates. We formally keep  the same  equation (\ref{eq:eigentwophoton0}), that is,
 \begin{equation}
\big(\beta_1 \bar{N}+ \beta_2 \bar{B}_-+\beta_3 \bar{B}_++\beta_4 \bar{A}_-+\beta_5 \bar{A}_+ \big)f(z)=\lambda f(z) \, ,
\label{eq:eigentwophoton}
\end{equation}
but now we  introduce the Dunkl--Fock--Bargmann representation (\ref{eq:reps2z}), which leads us to    the    differential-difference equation given by
 \begin{equation}
 \beta_2 \mathcal{D}_z^2 f(z)+(\beta_1 z+\beta_4)\mathcal{D}_z f(z)+\big(\beta_3 z^2+\beta_5 z-\mu \beta_1-\lambda \big)f(z)+\mu \beta_1 f(-z)=0 \, ,
 \label{eq:defeq}
 \end{equation}
which, in the limit $\mu \to 0$, reduces to the differential equation (\ref{eq:defeqmu01}).
Recall that   the $\beta_i$'s are arbitrary complex coefficients and $\lambda$ is a complex eigenvalue. 
Furthermore, it should be noted that, in principle, equation \eqref{eq:eigentwophoton} could also be extended  in the Dunkl framework by adding the reflection operator, namely
\begin{equation}
\big(\beta_1 \bar{N}+ \beta_2 \bar{B}_-+\beta_3 \bar{B}_++\beta_4 \bar{A}_-+\beta_5 \bar{A}_++\beta_6 \bar{R} \big)f(z)=\lambda f(z) \, ,
\nonumber
\end{equation}
  giving rise to the following differential-difference equation 
 \begin{equation}
\beta_2 \mathcal{D}_z^2 f(z)+(\beta_1 z+\beta_4)\mathcal{D}_z f(z)+\big(\beta_3 z^2+\beta_5 z-\mu \beta_1-\lambda \big)f(z)+(\mu \beta_1+\beta_6) f(-z)=0 \, ,
\label{eq:defeqR}
\end{equation}
which includes an additional ($\mu$-independent) constant $\beta_6$ in front of $f(-z)$. In this case, to recover equation (\ref{eq:defeqmu01}), we need to set $\beta_6=0$ and take the limit $\mu \to 0$. The physical significance of   equations (\ref{eq:defeq}) and (\ref{eq:defeqR})  in the context of  quantum optics remains an open problem.


\section{The Schr{\"o}dinger Lie algebra}
\label{sec4}

The results concerning the two-photon Lie algebra $\mathfrak{h}_6$, outlined  in  section~\ref{sec2}, can be translated into  a different mathematical and physical framework. The fundamental point is that the Lie algebra $\mathfrak{h}_6$ is isomorphic to the centrally extended Schr{\"o}dinger Lie algebra in $1+1$ dimensions, here denoted by  $\mathcal{S}$~\cite{PBH1997}. In this section, we review the (kinematical) structure of  $\mathcal{S}$ and its  role  as the symmetry algebra of the  $(1 + 1)$-dimensional  heat-Schr\"odinger equation (SE in short). These relations will be  extended  to the Dunkl context  in the next section, yielding several novel SE-type differential-difference equations.

The  (centrally extended) Schr\"odinger   algebra $\mathcal{S}$~\cite{Hagen,Niederer,Feinsilver2004} is   a six-dimensional Lie algebra, which can be    expressed in terms of the usual kinematical basis  $\{ D,\, P,\, K,\, H,\, C,\, M \}$ on the $(1+1)$-dimensional Galilean spacetime, such that the generator $D$ is a dilation, $P$ is a space translation,   $K$  is a Galilean boost, $H$ is a  time translation, 
 $C $ is  a conformal transformation  and the nontrivial central extension $M$ is the mass.
In this basis, the defining Lie brackets of  $\mathcal{S}$ are given by
  \begin{equation}
\begin{aligned}
 {[D,P]}&= -P \, , \quad  &[D, K] &= K \, ,   \quad &[D, H] &= -2H \, ,  \quad &[D, C] &= 2C \, ,  \\[2pt]
[K, H] &= P \, ,   \quad &[K, P] &= M \, ,  \quad &[P, H] &= 0 \, ,  \quad &[P, C] &= -K \, ,  \\[2pt]
 [K, C] &= 0 \, ,  \quad & [H, C] &= D  \, , \quad	& [M, \cdot\,] &= 0  \, .   &&
 \label{Dunkl-Schrodinger0}
\end{aligned}
\end{equation}
  
  Similarly to $\mathfrak{h}_6$, the    Schr\"odinger   algebra $\mathcal{S}$ covers some remarkable
   Lie subalgebras (see, e.g.,  \cite{PBH2000}):
 \begin{itemize}
\item The Heisenberg--Weyl    and the harmonic oscillator subalgebras:  $\mathfrak{h}_3={\rm span}\{ P,K,M \}$ and  $\mathfrak{h}_4={\rm span}\{ D,P,K,M \}$.
\item The $(1 + 1)$-dimensional extended Galilei algebra  $\bar{\mathcal{G}}={\rm span}\{  P,K,H,M \}$.
\item $\mathfrak{sl}(2,\mathbb R)={\rm span}\{D, H,C \}$ and $\mathfrak{gl}(2)={\rm span}\{D, H,C,M \}$; the latter is isomorphic to a direct sum of  $\mathfrak{sl}(2,\mathbb R)$ with the central extension $M$.
\end{itemize}
  Therefore, we find the following subalgebra embeddings:
\begin{equation}
\mathfrak{h}_3\subset \mathfrak{h}_4 \subset \mathcal{S}\,  ,\qquad  \mathfrak{h}_3\subset \bar{\mathcal{G}}\subset \mathcal{S}\,  ,\qquad
\mathfrak{sl}(2,\mathbb R) \subset\mathfrak{gl}(2)  \subset \mathcal{S}\,  .
\nonumber
\end{equation}

The map defined by
\begin{align}
\{ D,\, P,\, K,\, H,\, C,\, M  \} \to \{ -D,\, -K,\, -P,\, -C,\, -H,\, -M  \}
\label{autoS0}
\end{align}
is an involutive automorphism of $\mathcal{S}$, thus preserving the commutation relations (\ref{Dunkl-Schrodinger0}).

 In addition to $M$, the  Schr\"odinger   algebra $\mathcal{S}$  possesses  a nontrivial Casimir operator which can be computed either using the so-called ``matrix method" \cite{Campoamor2005} or the ``trace-formulae" approach \cite{Campoamor2020}. These procedures give rise to a fourth-order Casimir invariant which, upon factoring out an overall factor $M$, reduces to a cubic  Casimir invariant expressed as~\cite{CCH2025}
  \begin{align}
\CC_{\mathcal{S}} = 2 H C M-D P K-K^2H-P^2C-\frac{1}{2}D^2M-\frac{3}{2}D M \, ,
\label{eq:CasimirSx}
\end{align}
{\em i.e.},  $\big[\CC_{\mathcal{S}} , X \big] =0$ for all $X\in{\mathcal{S}} $.

  It is worth noting that there exists  a Lie algebra isomorphism between $\mathcal{S}$ and the two-photon algebra $ \mathfrak{h}_6$, namely~\cite{PBH1997}
\begin{equation}
D=-N-\frac12 M \, , \qquad P=A_+ \, , \qquad K=A_- \, , \qquad H=\frac 12 B_+  \, , \qquad C=\frac 12 B_-  \, ,
\label{changebasis0}
\end{equation}
keeping $M$ as the same central generator for both algebras.
It can be directly verified that this mapping
 relates the Lie brackets of  $ \mathfrak{h}_6$  (\ref{Two-Photon}) exactly to those of  $\mathcal{S}$ (\ref{Dunkl-Schrodinger0}), and vice versa.

Let us now outline the significance of  the Schr\"odinger   algebra $\mathcal{S}$ as the Lie symmetry algebra of the  SE. The kinematical nature of $\mathcal{S}$  arises in a natural way by recalling the usual differential realization of the Schr\"odinger generators, fuflilling (\ref{Dunkl-Schrodinger0}),  in terms of the space and time coordinates $(x, t)$  acting on a function $\ff(x,t)$:
\begin{equation}
\begin{aligned}
D&=2 t \partial_t+x \partial_x -a \, ,\qquad  P =\partial_x \, ,\qquad K=-t \partial_x -m x \, ,  \\[2pt]
H&=\partial_t \, , \qquad C=t^2 \partial_t+t x \partial_x +\frac{1}{2}mx^2-a t \, , \qquad M=m\, ,
\end{aligned}
\label{eq:DSH0}
 \end{equation}
  where $a$ and $m$ are the constants that label the representation, and from now on we will denote $\partial_x=\partial/{\partial x}$ and $\partial_t=\partial/{\partial t}$. Under this realization, the (free) SE 
can be straightforwardly obtained from  the Casimir operator of  the (extended) Galilei subalgebra  $\bar{\mathcal{G}}={\rm span}\{  P,K,H,M \}\subset \mathcal{S}$ given  by
\be
 \CC_{\bar{\mathcal{G}}}=P^2-2 H M \equiv E \, ,
\nonumber
\ee
 as follows
\begin{equation}
E \ff(x,t)=\big(P^2-2 H M\big)\ff(x,t)=0 \quad \Rightarrow \quad \big(\partial_x^2-2m \partial_t\big)\ff(x,t)=0 \, .
\label{eq:DSHE0}
\end{equation}
We will say that an operator $\OO=\OO(x,t,\partial_x,\partial_t)  $ is a symmetry of the linear differential equation $E \ff(x,t)=0$
if $\OO$ transforms solutions into solutions:
\be
(E\OO) \ff(x,t) = (\OO'E) \ff(x,t) \, ,
\label{symO}
\ee
where $\OO'=\OO'(x,t,\partial_x,\partial_t)$ is another operator. Therefore, since $E$ is provided by the Casimir operator $ \CC_{\bar{\mathcal{G}}}$ of the extended Galilei algebra, all its generators $\{  P,K,H,M \}$ are symmetries $\OO$ of the SE (\ref{eq:DSHE0}), trivially satisfying the relation (\ref{symO}) with $\OO'=\OO$.
Thus, in order to assert that the complete  Schr\"odinger   algebra $\mathcal{S}$  is a Lie symmetry algebra of the SE  (\ref{eq:DSHE0}), it is necessary to analyze the role of the two generators $D$ and $C$ outside of  $\bar{\mathcal{G}}$. According to the commutation rules (\ref{Dunkl-Schrodinger0}), the Lie bracket between $D$ and $ \CC_{\bar{\mathcal{G}}} \equiv E$ yields
\be
[E,D]=2E\quad \Rightarrow \quad (ED) \ff(x,t) = \big((D+2)E\big) \ff(x,t) \, ,
\label{ED}
\ee
so that $D$ is also a symmetry of the SE. Finally, we  compute the  Lie bracket using the conformal transformation and obtain that
\be
[E,C]= - (KP + PK + 2MD) \, .
\nonumber
\ee
In principle, it appears that $C$ is not a symmetry of the SE,  since this algebraic  result is not proportional to $E$. However, if take the particular representation (\ref{eq:DSH0}) with $a=-1/2$ (with $m$ arbitrary) we find that
\be
[E,C]=2t E\quad \Rightarrow \quad (EC) \ff(x,t) = \big((C+2t)E\big) \ff(x,t) \, ,
\label{EC}
\ee
proving the well-known result  that $\mathcal{S}$ is the Lie symmetry algebra of the SE  (\ref{eq:DSHE0}).


\section{The Dunkl--Schr{\"o}dinger algebra}
\label{sec5}

In this section, we develop the second objective of this paper in a ``parallel" manner as was already done for the  Dunkl two-photon algebra $\mathfrak{h}_6^{(\alpha)}$ in section~\ref{sec3}.
In particular, we present the  Dunkl--Schr{\"o}dinger algebra, denoted by $\mathcal{S}^{(\alpha)}$,  which requires considering the  reflection generator $R$. All   the corresponding algebraic relations for $\mathcal{S}^{(\alpha)}$ are shown in section~\ref{subsec5.1}, including the extension of the Lie algebra isomorphism  (\ref{changebasis0}) between  $\mathfrak{h}_6$ and $\mathcal{S}$ to the Dunkl framework.  Next, again from a purely algebraic point of view, we analyze the main (six) subalgebras of $\mathcal{S}^{(\alpha)}$,  together with their structural properties,   in section~\ref{subsec5.2}. Finally, new physical applications (completely different to those presented in sections~\ref{subsec3.3} and \ref{subsec3.4}) can be derived from the previous  algebraic structures for  $\mathcal{S}^{(\alpha)}$  by constructing   the  Dunkl representation counterpart of the standard differential realization (\ref{eq:DSH0}) of  $\mathcal{S}$, giving rise to new SE-type differential-difference equations.


\subsection{Definition, structural properties and Casimir invariant of $\mathcal{S}^{(\alpha)}$}
\label{subsec5.1}

The Dunkl--Schr{\"o}dinger algebra 
$\mathcal{S}^{(\alpha)}$ is spanned by   seven  generators 
$\{ D,\, P,\, K,\, H,\, C,\, M,\, R \} $, which we will assume to be ordered in the form 
$$
D  \prec P \prec K  \prec H  \prec C \prec M  \prec R \, .
$$
The defining commutation rules of $\mathcal{S}^{(\alpha)}$ are given by
\begin{equation}
\begin{aligned}
 {[D,P]}&= -P \, , \quad  &[D, K] &= K \, ,   \quad &[D, H] &= -2H \, ,  \quad &[D, C] &= 2C \, ,  \\[2pt]
[K, H] &= P \, ,   \quad &[K, P] &= M (\mathrm{Id} + \alpha R) \, ,  \quad &[P, H] &= 0 \, ,  \quad &[P, C] &= -K \, ,  \\[2pt]
 [K, C] &= 0 \, ,  \quad & [H, C] &= D  \, , \quad	& [M, \cdot\,] &= 0  \, ,   &&
 \label{Dunkl-Schrodinger}
\end{aligned}
\end{equation}
where $\alpha \in \mathbb{R}$ and $\mathrm{Id}$ is the identity element,  
  together with the   additional relations involving the reflection generator $R$:
  \begin{equation}
\begin{aligned}
R D &= D R  \, , \quad  &R P &= -P R \, ,   \quad &R K &= - K R  \, ,    \\[2pt]
R H &=  H R \, ,   \quad &R C &= C R \, ,  \quad  &R M &= M R \, ,  \label{Dunkl-Schrodinger2}
\end{aligned}
\end{equation}
such that $R^2 = \mathrm{Id}$.

 Notice also that  $\mathcal{S}^{(\alpha)}$ is endowed with the involutive automorphism defined by
\begin{align}
\{ D,\, P,\, K,\, H,\, C,\, M,\, R \} \to \{ -D,\, -K,\, -P,\, -C,\, -H,\, -M,\, R \}
\nonumber
\end{align}
 that preserves the  commutations relations  (\ref{Dunkl-Schrodinger}) and (\ref{Dunkl-Schrodinger2}), thus extending the map (\ref{autoS0}) to the Dunkl setting.

 The relationship between $\mathcal{S}^{(\alpha)}$ and the Dunkl two-photon algebra $\mathfrak{h}_6^{(\alpha)}$ is established via  
 the following change of basis
\begin{equation}
D=-N-\frac12(1+\alpha)M \, , \qquad P=A_+ \, , \qquad K=A_- \, , \qquad H=\frac 12 B_+  \, , \qquad C=\frac 12 B_-  \, ,
\label{changebasis}
\end{equation}
keeping the central generator $M$ and the reflection generator $R$ for both $\mathcal{S}^{(\alpha)}$ and $\mathfrak{h}_6^{(\alpha)}$. This  map generalizes the Lie algebra isomorphism (\ref{changebasis0}) and exactly relates the commutation relations (\ref{Dunkl Two-Photon}) and (\ref{addrel}) 
 for $\mathfrak{h}_6^{(\alpha)}$ to the corresponding ones (\ref{Dunkl-Schrodinger}) and (\ref{Dunkl-Schrodinger2}) for $\mathcal{S}^{(\alpha)}$.

Again, the reflection generator $R$ induces a $\mathbb{Z}_2$ grading of the underlying vector space of $\mathcal{S}^{(\alpha)}$:
\begin{equation*}
\mathcal{S}^{(\alpha)} =\mathcal{S}_{0}^{(\alpha)}  \oplus \mathcal{S}_{1}^{(\alpha)} \, ,
\end{equation*}
where 
\begin{equation*}
\mathcal{S}_{0}^{(\alpha)} := \mathrm{span}\{ D, H, C, M, R \} \, ,
\qquad 
\mathcal{S}_{1}^{(\alpha)} := \mathrm{span}\{ P,K \} \, ,
\end{equation*}
and
\begin{equation*}
R X = (-1)^{|X|} X R\, , 
\quad \text{with}\quad  |X| = 
\begin{cases}
0\, , & X \in \mathcal{S}_{0}^{(\alpha)} \, ,\\[4pt]
1\, , & X \in \mathcal{S}_{1}^{(\alpha)} \, .
\end{cases}
\end{equation*}
Therefore, the  reflection generator  $R$  distinguishes between even and odd generators, as expected.

As in remark \ref{rem1}, we would like to add the following comments.

\begin{rem}
The structure $\big(\mathcal{S}^{(\alpha)} , \boldsymbol{\cdot}\,, \mathrm{Id}\big)$ is an associative unital noncommutative algebra. The commutator
\[
[X,Y] := XY - YX\, , \qquad X,Y \in \mathcal{S}^{(\alpha)} \, ,
\]
provides $\mathcal{S}^{(\alpha)}$ with the structure of a generally nonlinear algebra.
In addition, since $\mathcal{S}^{(\alpha)}$ is associative, the commutator automatically satisfies the
Jacobi identity,
\[
[X,[Y,Z]] + [Y,[Z,X]] + [Z,[X,Y]] = 0\, ,
\]
for all $X,Y,Z \in \mathcal{S}^{(\alpha)}$. When the reflection operator is discarded  and the  limit $\alpha \to 0$ is considered, the Dunkl Schr{\"o}dinger algebra $\mathcal{S}^{(\alpha)}$ reduces to the standard centrally extended Schr\"odinger Lie algebra in $(1+1)$ dimensions $\mathcal{S}$ described in section~\ref{sec4}.
\end{rem}

Finally, it  can be shown that the (nontrivial) Casimir invariant of the Dunkl--Schr{\"o}dinger algebra $\mathcal{S}^{(\alpha)}$  can be expressed as the   cubic polynomial given by
  \begin{align}
\CC_{\mathcal{S}^{(\alpha)}} = 2 H C M-D P K-K^2H-P^2C-\frac{1}{2}D^2M-\frac{3}{2}D M-\frac{\alpha}{2}\left(D M R-\frac{1}{2} M R \right) \, ,
\label{eq:CasimirS}
\end{align}
which consistently  reduces to  the expression $\CC_{\mathcal{S}}$ (\ref{eq:CasimirSx}) for   $\mathcal{S}$.


\subsection{$\mathcal{S}^{(\alpha)}$-subalgebras and corresponding Casimir invariants}
\label{subsec5.2}

 Following the same line of reasoning developed in section~\ref{subsec3.2} regarding the Dunkl two-photon algebra  $\mathfrak{h}_6^{(\alpha)}$, we present  the same six relevant five-dimensional subalgebras of $\mathcal{S}^{(\alpha)}$.    We use the same terminology as in section~\ref{subsec3.2} and display  their main algebraic structure in Table~\ref{table2}, which is related to Table~\ref{table1}
through the map (\ref{changebasis}).
 
We   briefly discuss the results in Table~\ref{table2}. The four-dimensional   Dunkl Heisenberg--Weyl algebra 
 in the $\mathcal{S}^{(\alpha)}$-basis corresponds to
    \be
  \mathfrak{h}_3^{(\alpha)}={\rm span}\{ P,K,M,R  \}  
\nonumber
  \ee
  with commutation relations now given by
 \be
  [K, P] = M (\mathrm{Id} + \alpha R) \, , \qquad R P=-P R \, ,\qquad R K=-K R  \, , \qquad  [M, \cdot\, ]  = 0  \, .
\nonumber
 \ee
And the same  $\mathfrak{h}_6^{(\alpha)}$-subalgebra embeddings (\ref{embeddings}) also hold for $\mathcal{S}^{(\alpha)}$. In the kinematical basis of $\mathcal{S}^{(\alpha)}$, the proper Dunkl extended Galilei subalgebra 
corresponds to $\bar{\mathcal{G}}_+^{(\alpha)}={\rm span}\{  P,K,H,M,R  \}$, although there exists another Galilei subalgebra $\bar{\mathcal{G}}_-$. Similarly to Table~\ref{table1}, note that we  have defined a new rotation generator $ \JJ_+$ for the Dunkl 
extended Euclidean subalgebra $\bar{\mathcal{E}}^{(\alpha)}$ and a new boost generator $ \JJ_-$ for the Dunkl 
extended Poincar\'e subalgebra $\bar{\mathcal{P}}^{(\alpha)}$. The set of generators $\{D, \JJ_\pm, M,R\}$ span a 
    Dunkl subalgebra isomorphic to  $\mathfrak{gl}(2)^{(\alpha)}$ with non-vanishing commutation rules reading as 
  \begin{equation*} 
  [D,\JJ_+]=2\JJ_-\, ,\qquad [D,\JJ_-]=2\JJ_+\,  , \qquad   [\JJ_-,\JJ_+]= -2D\, .
\end{equation*}

  We calculate the commutators between the two generators outside of   each of
the first five subalgebras  in Table~\ref{table2} finding that 
  \begin{equation}
\begin{aligned}
    \mathfrak{h}_4^{(\alpha)}&\!: \quad & \big[\CC_{\mathfrak{h}_4^{(\alpha)}} ,H\big] &=   \CC_{ \bar{\mathcal{G}}_+^{(\alpha)} } \, , \quad &  \big[\CC_{\mathfrak{h}_4^{(\alpha)}} ,C\big] &=- \CC_{\bar{\mathcal{G}}_-^{(\alpha)} } \, .\\[1pt]
     \bar{\mathcal{G}}_+^{(\alpha)}&\!: \quad    & \big[\CC_{ \bar{\mathcal{G}}_+^{(\alpha)}  } ,D\big] &=  2\CC_{ \bar{\mathcal{G}}_+^{(\alpha)} } \, , \quad &  \big[\CC_{ \bar{\mathcal{G}}_+^{(\alpha)} } ,C\big] &= -2 \CC_{\mathfrak{h}_4^{(\alpha)}}  -  M   \, .\\[1pt]
      \bar{\mathcal{G}}_-^{(\alpha)} &\!: \quad    & \big[\CC_{ \bar{\mathcal{G}}_-^{(\alpha)}  } ,D\big] &=  -2 \CC_{\bar{\mathcal{G}}_-^{(\alpha)} } \, , \quad &  \big[\CC_{ \bar{\mathcal{G}}_-^{(\alpha)} } ,H\big] &= 2 \CC_{\mathfrak{h}_4^{(\alpha)}}  +  M   \, .\\[1pt]
     \bar{\mathcal{E}}^{(\alpha)}&\!: \quad    & \big[\CC_{ \bar{\mathcal{E}}^{(\alpha)} } ,D\big] &= - 2 \CC_{ \bar{\mathcal{P}}^{(\alpha)} } \, , \quad &  \big[\CC_{ \bar{\mathcal{E}}^{(\alpha)} } ,\JJ_-\big] &=  -4 \CC_{\mathfrak{h}_4^{(\alpha)}}  - 2 M   \, .\\[1pt]
          \bar{\mathcal{P}}^{(\alpha)}&\!: \quad    & \big[\CC_{ \bar{\mathcal{P}}^{(\alpha)} } ,D\big] &=-  2 \CC_{ \bar{\mathcal{E}}^{(\alpha)} } \, , \quad &  \big[\CC_{ \bar{\mathcal{P}}^{(\alpha)} } ,\JJ_+\big] &=   4 \CC_{\mathfrak{h}_4^{(\alpha)}}  + 2 M   \, .
\nonumber
            \end{aligned}
\end{equation}
To be compared with (\ref{comcash6}).   However,  for ${\mathfrak{gl}(2)^{(\alpha)} } $ we obtain commutators that do not involve other Casimir invariants (see (\ref{comcash6b})):
\begin{equation}
\begin{aligned}
\big[\CC_{\mathfrak{gl}(2)^{(\alpha)} } ,P\big] &= 2 DP -P + 4 KH   \, , \\[4pt]
\big[\CC_{\mathfrak{gl}(2)^{(\alpha)} } ,K\big] &= -2 DK-K -4 PC  \, . 
  \end{aligned}
\nonumber
\end{equation}

  There is a relationship between the cubic Casimir $\CC_{\mathcal{S}^{(\alpha)}} $  of the algebra $\mathcal{S}^{(\alpha)}$ (\ref{eq:CasimirS}) with the Casimir invariants of the three subalgebras $\bar{\mathcal{G}}_\pm^{(\alpha)}$ and $\mathfrak{h}_4^{(\alpha)}$, and the  central generator $M$, which turns out to be  
  \begin{equation}
\frac{1}{2}\left\{\CC_{\bar{\mathcal{G}}_+^{(\alpha)}} ,\CC_{\bar{\mathcal{G}}_-^{(\alpha)}}\right\}  -\left(\CC_{\mathfrak{h}_4^{(\alpha)}}\right)^2-\CC_{\mathfrak{h}_4^{(\alpha)}} M-\left(1-\frac{\alpha^2}{4}\right)M^2=2 \CC_{\mathcal{S}^{(\alpha)}} M \, ,
\nonumber
\end{equation}
 where again $\{\cdot, \cdot\}$ denotes the anticommutator. If $R$ is cancelled with $\alpha=0$ and the global central term $M$ is removed, then we recover the cubic  Casimir invariant $\CC_{\mathcal{S}}$ (\ref{eq:CasimirSx}) for   $\mathcal{S}$.


 \begin{table}[t!]
{ 
\caption{  {Six relevant  five-dimensional   subalgebras of the Dunkl--Schr\"odinger algebra  $\mathcal{S}^{(\alpha)}$ with commutations relations given by (\ref{Dunkl-Schrodinger}) and (\ref{Dunkl-Schrodinger2}). For each case, the non-vanishing commutation rules and the (quadratic) Casimir polynomial are given.}}

\begin{center} 
\begin{tabular}{l}
\hline

\hline
\\[-4pt]
$\bullet$ Dunkl harmonic oscillator subalgebra: $\mathfrak{h}_4^{(\alpha)}={\rm span}\{ D,P,K,M,R  \}$ \\[4pt]
\phantom{$\bullet$}   $[D, P] = -P\qquad [D, K] = K\qquad  [K, P] = M (\mathrm{Id} + \alpha R)\qquad R P=-P R \qquad R K=-K R $
\\[4pt]
\phantom{$\bullet$}   $\CC_{\mathfrak{h}_4^{(\alpha)}} =DM + PK +\frac{\alpha}{2}MR$
   \\[8pt]
   
$\bullet$ Dunkl centrally extended Galilei subalgebra: $\bar{\mathcal{G}}_+^{(\alpha)}={\rm span}\{  P,K,H,M,R  \}$ \\[4pt]
\phantom{$\bullet$}   $[K, H]  = P \qquad   \ \  [K, P] = M (\mathrm{Id} + \alpha R)\qquad R P=-P R \qquad R K=-K R$
\\[4pt]
\phantom{$\bullet$}   $\CC_{\bar{\mathcal{G}}_+^{(\alpha)} } =P^2-2HM $
   \\[8pt]

$\bullet$ Dunkl centrally extended Galilei subalgebra: $\bar{\mathcal{G}}_-^{(\alpha)}={\rm span}\{  P,K,C,M,R  \}$ \\[4pt]
\phantom{$\bullet$}   $[P, C]  = -K \qquad  [K, P] = M (\mathrm{Id} + \alpha R)\qquad R P=-P R \qquad R K=-K R$
\\[4pt]
\phantom{$\bullet$}   $\CC_{\bar{\mathcal{G}}_-^{(\alpha)} } =K^2-2 CM $
   \\[8pt]

$\bullet$ Dunkl centrally extended Euclidean subalgebra: $\bar{\mathcal{E}}^{(\alpha)}={\rm span}\big\{ \JJ_+:=   C+H , P,K, M,R \big \}$ \\[4pt]
\phantom{$\bullet$}   $[\JJ_+, P]  = K  \qquad [\JJ_+, K]  =  -P \qquad  [K, P] = M (\mathrm{Id} + \alpha R)\qquad R P=-P R \qquad R K=-K R $
\\[4pt]
\phantom{$\bullet$}   $\CC_{ \bar{\mathcal{E}}^{(\alpha)} }= K^2 +P^2-  2  \JJ_+ M $
   \\[8pt]

$\bullet$ Dunkl centrally extended Poincar\'e subalgebra: $\bar{\mathcal{P}}^{(\alpha)}={\rm span}\big\{ \JJ_-:=   C-H , P,K, M,R \big \}$ \\[4pt]
\phantom{$\bullet$}   $[ \JJ_-, P]  = K  \qquad [ \JJ_-, K]  =   P \qquad\phantom{-}  [K, P] = M (\mathrm{Id} + \alpha R)\qquad R P=-P R \qquad R K=-K R $
\\[4pt]
\phantom{$\bullet$}   $\CC_{ \bar{\mathcal{E}}^{(\alpha)} }= K^2 -P^2-  2   \JJ_- M $
   \\[8pt]
   
$\bullet$ Dunkl  $\mathfrak{gl}(2) $ subalgebra: $\mathfrak{gl}(2)^{(\alpha)}={\rm span}\{ D,  H,  C, M,R  \}$ \\[4pt]
\phantom{$\bullet$}   $[D, H]  =-2 H \qquad [D, C]  = 2C\qquad [H, C]  = D $
\\[4pt]
\phantom{$\bullet$}   $\CC_{\mathfrak{gl}(2)^{(\alpha)} } =  4 HC - D^2 - 2 D $
     \\[8pt]
 \hline
				
\hline
\end{tabular}
\end{center} 
\label{table2}
}
\end{table}



\subsection{Vector field realization   and Dunkl heat-Schr{\"o}dinger type equations}
\label{subsec5.3}
  
Once  the Dunkl--Schr{\"o}dinger algebra $\mathcal{S}^{(\alpha)}$ has been constructed and  its main algebraic properties derived, we can apply them to obtain new differential-difference equations, which   include, among others, the Dunkl counterpart of the SE (\ref{eq:DSHE0}). 

The cornerstone of our approach  requires, first and foremost,   deducing  a differential-difference representation for $\mathcal{S}^{(\alpha)}$ that satisfies the commutation relations (\ref{Dunkl-Schrodinger}) and (\ref{Dunkl-Schrodinger2}). Provided that the parameter $\alpha=2 \mu$, the Dunkl vector field realization of $\mathcal{S}^{(\alpha)}$ in terms of the space and time coordinates $(x, t)$  and acting on a function $\ff(x,t)$ is found to be
  \begin{equation}
\begin{aligned}
D&=2 t \partial_t+x \mathcal{D}_x+\mu R-a \, ,\qquad P=\mathcal{D}_x \, ,\qquad K=-t \mathcal{D}_x-m x \, ,  \\[2pt]
H&=\partial_t \, , \qquad C=t^2 \partial_t+t x \mathcal{D}_x+\mu t R+\frac{1}{2}mx^2-a t \, , \qquad M=m\, ,
\label{Dunkl-Schrodinger2x}
\end{aligned}
\end{equation}
together with the usual spatial action of the reflection generator $R \ff(x,t)=\ff(-x,t)$, and where now $a$, $m$  and $\mu$ are arbitrary real constants. Consequently, this realization keeps the time coordinate $t$ continuous, while the Dunkl derivative $ \mathcal{D}_x$, defined in (\ref{Dderivative}), affects the space coordinate $x$.
 The value of the cubic Casimir of $\mathcal{S}^{(\alpha)}$ (\ref{eq:CasimirS})  through the above realization gives a constant:
 \begin{equation}
C_{\mathcal{S}^{(\alpha)}}=\frac{1}{2}m\big(\mu^2-a(a+3)-2\big) \, .
\nonumber
\end{equation}
Note that the rotation/boost generator $\JJ_\pm$ for $\bar{\mathcal{E}}^{(\alpha)}$ and $\bar{\mathcal{P}}^{(\alpha)}$ in Table~\ref{table2} adopt the following realization
\be
\JJ_\pm=C\pm H= \big( t^2 \pm 1\big)\partial_t+t x \mathcal{D}_x+\mu t R+\frac{1}{2}mx^2-a t \, .
\label{reJpm}
\ee

Now we constuct the Dunkl SE following the procedure described in section~\ref{sec4}. We start from the
Dunkl centrally extended Galilei subalgebra $\bar{\mathcal{G}}_+^{(\alpha)}={\rm span}\{  P,K,H,M,R  \}$ in Table~\ref{table2} and define the operator $E$ from its Casimir invariant as in (\ref{eq:DSHE0}):
\be
\CC_{\bar{\mathcal{G}}_+^{(\alpha)} } =P^2-2 H M \equiv E \, .
 \label{casG1}
\ee
Hence, the Dunkl SE reads:
\begin{equation}
E \ff(x,t)=\big(P^2-2 H M\big)\ff(x,t)=0 \quad \Rightarrow\quad  \big(\mathcal{D}^2_x-2m \partial_t\big)\ff(x,t)=0 \, .
\label{eq:DSH}
\end{equation}
Therefore, the five operators $\{  P,K,H,M,R  \}$ are automatically symmetries of the equation since they commute with $E$, {\em i.e.} they transform solutions into solutions (\ref{symO}). For the two remaining $\mathcal{S}^{(\alpha)}$ generators outside $\bar{\mathcal{G}}_+^{(\alpha)}$, it is found that they are symmetries provided that $a=-1/2$:
\be
[E,D]=2E\, ,\qquad [E,C]=2t E\, ,
\nonumber
\ee
as in relations (\ref{ED}) and (\ref{EC}). Hence, the complete Dunkl--Schr{\"o}dinger algebra $\mathcal{S}^{(\alpha)}$ provides seven symmetries of the Dunkl SE (\ref{eq:DSH}).

 Subsequently, we derive and analyze the corresponding  differential-difference equations arising from the five remaning $\mathcal{S}^{(\alpha)}$ subalgebras in Table~\ref{table2}, starting from their Casimir invariant.

 
 \noindent
 $\bullet$
 {\em Dunkl harmonic oscillator subalgebra} $\mathfrak{h}_4^{(\alpha)}={\rm span}\{ D,P,K,M,R  \}$.\\[2pt]
The associated differential-difference equation is obtained by introducing the realization (\ref{Dunkl-Schrodinger2x}) into the Casimir of  $\mathfrak{h}_4^{(\alpha)}$ leading to  
\begin{equation}
C_{\mathfrak{h}_4^{(\alpha)}}=-t E-m(1+a) \, ,
\nonumber
\end{equation}
where the operator $E\equiv \CC_{\bar{\mathcal{G}}_+^{(\alpha)} } $ is written in (\ref{casG1}).
Based on this result, we find a differential-difference equation endowed with five symmetries provided by $\mathfrak{h}_4^{(\alpha)}$:
\begin{equation}
C_{\mathfrak{h}_4^{(\alpha)}} \ff(x,t)=- \big(t E+m(1+a)\big)\ff(x,t)=0 \quad \Rightarrow\quad  \big( t \mathcal{D}^2_x-2m t \partial_t + m(1+a)\big)\ff(x,t)=0 \, .
\label{eq:DSHo}
\end{equation}
Next, we study if the two remaining generators $H$ and $C$ are additional symmetries of (\ref{eq:DSHo}) by imposing that they also transform solutions into solutions (\ref{symO}).
For each operator we obtain a  different value for the parameter $a$, which means that we arrive at a particular equation (\ref{eq:DSHo}) with {\em six} symmetries:
    \begin{equation}
\begin{aligned}
a&= -1:   \quad  &\big[C_{\mathfrak{h}_4^{(\alpha)}}, H \big ] &= -t^{-1} C_{\mathfrak{h}_4^{(\alpha)}} \, ,   \quad & t\big(   \mathcal{D}^2_x-2m  \partial_t  \big)\ff(x,t)&=0   \, .   \\[2pt]
a &= 0:  \quad    &\big[C_{\mathfrak{h}_4^{(\alpha)}}, C \big ]  &= t  C_{\mathfrak{h}_4^{(\alpha)}}  \, ,  \quad  &\big(   t\mathcal{D}^2_x-2m  t\partial_t +m \big)\ff(x,t) &=0\, .
\nonumber
\end{aligned}
\end{equation}

 \noindent
 $\bullet$
{\em Dunkl centrally extended Galilei subalgebra} $\bar{\mathcal{G}}_-^{(\alpha)}={\rm span}\{  P,K,C,M,R  \}$.\\[2pt]
We now consider the second Dunkl extended Galilei subalgebra. The image of its Casimir, under the realization (\ref{Dunkl-Schrodinger2x}),  gives
\begin{equation}
C_{\bar{\mathcal{G}}_-^{(\alpha)}}=t^2 E +m(1+2 a)t \, ,
\nonumber
\end{equation}
yielding the differential-difference equation  given by
\begin{equation}
C_{\bar{\mathcal{G}}_-^{(\alpha)}}\ff(x,t)=  \big( t^2 E +m(1+2 a)t \big)\ff(x,t)=0 \quad \Rightarrow\quad  \big( t^2 \mathcal{D}^2_x-2m t^2 \partial_t + m(1+2a)t\big)\ff(x,t)=0 \, .
\label{eq:DSGm}
\end{equation}
The dilation operator $D$ is also a symmetry of this equation since 
$$
\big[C_{\bar{\mathcal{G}}_-^{(\alpha)}} ,D\big] = -2 C_{\bar{\mathcal{G}}_-^{(\alpha)}}\, ,
$$ 
without any restriction on $a$. Therefore, the equation (\ref{eq:DSGm}) is always endowed with six symmetries. 
 Nevertheless, for the operator $H$ it is necessary to fix $a=-1/2$, obtaining a particular case of (\ref{eq:DSGm}) with {\em seven} symmetries, expressed explicitly as
\be
a=-1/2:\qquad \big[C_{\bar{\mathcal{G}}_-^{(\alpha)}} ,H\big] =- 2 t^{-1} C_{\bar{\mathcal{G}}_-^{(\alpha)}}\, ,\qquad  t^2\big(   \mathcal{D}^2_x-2m  \partial_t  \big)\ff(x,t) =0\, .
\nonumber
\ee

 \noindent
 $\bullet$
{\em Dunkl centrally extended Euclidean subalgebra} $\bar{\mathcal{E}}^{(\alpha)}={\rm span}\big\{ \JJ_+ , P,K, M,R \big \}$.\\[2pt]
In this case we consider the realization (\ref{Dunkl-Schrodinger2x})  together with (\ref{reJpm}), obtaining that
\begin{equation}
C_{\bar{\mathcal{E}}^{(\alpha)}}=(1+t^2) E +m(1+2 a)t \, ,
\nonumber
\end{equation}
so that the associated differential-difference equation turns out to be
\be
\left( (1+t^2) \big(  \mathcal{D}^2_x-2m   \partial_t \big) + m(1+2a)t  \right)\ff(x,t)=0\, .
\label{eq:DSEuc}
\ee
The generators $D$ and $\JJ_-$ and symmetries of the particular equation with $a=-1/2$ since they satisfy the following commutators with the Euclidean Casimir operator:
\be
\big[\CC_{ \bar{\mathcal{E}}^{(\alpha)} } ,D\big] = -2\, \frac{t^2-1}{t^2+1} \,\CC_{ \bar{\mathcal{E}}^{(\alpha)} }  \, , \qquad \big[\CC_{ \bar{\mathcal{E}}^{(\alpha)} } ,\JJ_-\big] =   \frac{4 t}{t^2+1} \,\CC_{ \bar{\mathcal{E}}^{(\alpha)} }  \, .
\nonumber
\ee
Consequently, the equation (\ref{eq:DSEuc}) has always five symmetries, while the particular case
\be
  (1+t^2) \big(  \mathcal{D}^2_x-2m   \partial_t \big)   \ff(x,t)=0\, ,
  \nonumber
\ee
is endowed with seven symmetry operators.

 \noindent
 $\bullet$
{\em Dunkl centrally extended Poincar\'e subalgebra} $\bar{\mathcal{P}}^{(\alpha)}={\rm span}\big\{ \JJ_-  , P,K, M,R \big \}$. \\[2pt]
Now we obtain that
\begin{equation}
C_{\bar{\mathcal{P}}^{(\alpha)}}=(t^2-1) E +m(1+2 a)t \, ,
\nonumber
\end{equation}
which gives rise to the differential-difference equation given by
\be
\left( ( t^2-1) \big(  \mathcal{D}^2_x-2m   \partial_t \big) + m(1+2a)t \right)\ff(x,t)=0\, .
\label{eq:DSEucp}
\ee
Again, the two remaining generators,  $D$ and $\JJ_+$, are also symmetries provided that $a=-1/2$:
\be
\big[\CC_{ \bar{\mathcal{P}}^{(\alpha)} } ,D\big] = -2\, \frac{t^2+1}{t^2-1} \,\CC_{ \bar{\mathcal{P}}^{(\alpha)} }  \, , \qquad \big[\CC_{ \bar{\mathcal{P}}^{(\alpha)} } ,\JJ_+\big] =  - \frac{4 t}{t^2-1} \,\CC_{ \bar{\mathcal{P}}^{(\alpha)} }  \, .
\nonumber
\ee
And the particular case of  (\ref{eq:DSEucp}) with $a=-1/2$ possesses seven symmetry operators:
\be
  ( t^2-1) \big(  \mathcal{D}^2_x-2m   \partial_t \big)  \ff(x,t)=0\, .
  \nonumber
\ee


 \noindent
 $\bullet$
{\em Dunkl $\mathfrak{gl}(2)$ subalgebra} $\mathfrak{gl}(2)^{(\alpha)}={\rm span}\{ D,  H,  C, M,R  \}$. \\[2pt]
We construct the final  differential-difference equation starting from $\mathfrak{gl}(2)^{(\alpha)}$.
 The Casimir operator adopts the following expression
 $$
\CC_{\mathfrak{gl}(2)^{(\alpha)}}=- x^2 E  +(1+2a )x\mathcal{D}_x  + 2 (1+a)\mu R- a(2+a)-\mu^2 \, ,
$$
thus leading to the corresponding differential-difference equation, namely
\be
\bigg( -x^2 \mathcal{D}_x^2 + 2 m x^2 \partial_t+(1+2a )x\mathcal{D}_x + 2 (1+a)\mu R- a(2+a)-\mu^2
  \bigg)\ff(x,t)=0\, .
\nonumber
\ee
Unlike in the previous cases, if we assume that the Dunkl structure (with $\mu\ne 0$) is preserved,   it can be shown that there are no additional symmetries arising from $\mathcal{S}^{(\alpha)}$ for the above equation, regardless of the value of $a$. This statement follows from the commutators between $P$ and $K$ with the Casimir of $\mathfrak{gl}(2)^{(\alpha)}$:
 \begin{equation}
\begin{aligned}
\big[\CC_{\mathfrak{gl}(2)^{(\alpha)} } ,P\big] &=  2 x \mathcal{D}_x^2-(1+2a-2\mu R)\mathcal{D}_x-4 m x \partial_t      \, , \\[2pt]
\big[\CC_{\mathfrak{gl}(2)^{(\alpha)} } ,K\big] &= -2 t x \mathcal{D}_x^2+(1+2 a-2\mu R)t\mathcal{D}_x+4 m t x\partial_t -m x(1+2a-2  \mu  R) \, . 
  \end{aligned}
  \nonumber
\end{equation}


\section{Concluding remarks}
\label{conc}

In this work, we have presented the novel  Dunkl two-photon and Schr{\"o}dinger algebras within a purely algebraic framework and derived their corresponding cubic Casimir invariants. We have also identified and studied for both algebras six relevant Dunkl subalgebras, together with their associated quadratic Casimir invariants. Furthermore, we have provided explicit realizations and applications of these algebraic structures. On the one hand, for the Dunkl two-photon algebra, we have constructed representations on the Fock space of the one-dimensional Dunkl harmonic oscillator and, subsequently, obtained the holomorphic realization in the associated Dunkl--Fock--Bargmann space.
On the other hand, for the Dunkl--Schr{\"o}dinger algebra, we have deduced an explicit vector field realization involving the Dunkl spatial derivative while keeping the time variable continuous. When applied to the Casimir invariants of the aforementioned subalgebras, this realization has given rise to    six new families of   Dunkl differential-difference equations in $1+1$ dimensions, including, in particular, a Dunkl counterpart of the well-known heat-Schr{\"o}dinger equation.

There are some open problems that emerge from our results which may deserve a more profound investigation. Specifically:
\begin{itemize}

\item 
The Dunkl approach requires to  introduce a  reflection operator $R$ determined by a parameter,  denoted here by  $\alpha$ and for representations  $\alpha=2\mu>-1$. This extension leads to  differential-difference representations  in one variable from an initial given Lie algebra, in this work the spatial one. The quantum group formalism (Hopf algebra deformations) for a Lie algebra also yields 
a discretization on one variable (see~\cite{BHNN2000} and references therein),  although in this case it is determined by the quantum deformation parameter.  Therefore, we consider that it would be interesting to develop a formalism unifiying  Dunkl and quantum groups approaches.

\item  In line with the ideas set out above, even without a quantum group deformation (which, in principle, at a first step would  not be necessary), a Hopf algebra structure for a Lie algebra ``extended" by a reflection operator could give rise to new systems and applications in arbitrary dimensions through the coalgebra approach~\cite{BBHMR2009,HL2025, GLG2026}, which would include  other Dunkl  algebras beyond the two-photon and Schr{\"o}dinger algebras  studied here.

\item  Finally, Lie and  Lie--Hamilton systems~\cite{LS20}  are systems of first-order   ordinary differential equations  determined by some $t$-dependent functions which are endowed with a superposition rule. The role of Dunkl operators, coalgebra symmetry and   Hopf algebra deformations, to the best of our knowledge,  has not yet been adequately formalized within  a unified theory.

\end{itemize}

Work along these  lines is currently in progress.


\section*{Acknowledgements}

\phantomsection
\addcontentsline{toc}{section}{Acknowledgements}

\small
  F.J.H. has been partially supported by Agencia Estatal de Investigaci\'on (Spain) under  the grant PID2023-148373NB-I00 funded by MCIN/AEI/10.13039/501100011033/FEDER, UE. F.J.H. also acknowledges support by the Regional Government of Castilla y Le\'on (Junta de Castilla y Le\'on, Spain) under the grant BU011P25/FEDER, UE.  D.L. is supported by HORIZON EUROPE - European Research Council (ERC) - STARTING GRANT 2021 ``Hamiltonian Dynamics, Normal Forms and Water Waves" (HamDyWWa) - Project Number: HE$\_$ERC22RMONT$\_$01.
  Views and opinions expressed are however those of the author only and do not necessarily reflect those
  of the European Union or the European Research Council. Neither the European Union nor the granting authority can be held responsible for them. D.L. is also partially supported by INFN-CSN4 (Commissione Scientifica Nazionale 4 - Fisica Teorica), MMNLP project. D.L. is a member of GNFM, INdAM.

\section*{Conflict of interest}
\phantomsection
\addcontentsline{toc}{section}{Conflict of interest}
The authors declare no conflicts of interest.

\end{document}